\documentclass[iop]{emulateapj}

\newcommand{\TVLM}{TVLM~0513-46}
\newcommand{\MASS}{2M~0746+20}
\newcommand{\vsini}{$v$~sin~$i$ }

\usepackage{amsmath}
\usepackage[caption=false]{subfig}
\usepackage{natbib}
\usepackage{graphicx}
\usepackage{longtable}

\makeatletter
\def\LT@makecaption#1#2#3{%
  \LT@mcol\LT@cols c{\hbox to\z@{\hss\parbox[t]\LTcapwidth{%
    \sbox\@tempboxa{#1{#2 }#3}%
    \ifdim\wd\@tempboxa>\hsize
      #1{#2 }#3%
    \else
      \hbox to\hsize{\hfil\box\@tempboxa\hfil}%
    \fi
    \endgraf\vskip\baselineskip}%
  \hss}}}
\makeatother

\begin{document}
\title{Wideband Dynamic Radio Spectra of Two Ultra-cool dwarfs}

\author{C. Lynch$^1$, R. L. Mutel$^1$, M. G\"udel$^2$}
\affil{$^1$Department of Physics and Astronomy, University of Iowa, Iowa City, Iowa 52240, USA}
\affil{$^2$Department of Astrophysics, University of Vienna, Vienna, AT}
\begin{abstract}

A number of radio-loud ultra cool dwarf stars (UCD) exhibit both continuous broadband and highly polarized pulsed radio emission. In order to determine the nature of the emission and the physical characteristics in the source region, we have made multi-epoch, wideband spectral observations of \TVLM\ and \MASS. We combine these observations with archival radio data to fully characterize both the temporal and spectral properties of the radio emission.  The continuum spectral energy distribution can be well modeled using gyrosynchrotron emission from mildly relativistic electrons in a dipolar field. The pulsed emission exhibits a variety of time-variable characteristics, including frequency drifts, frequency cutoffs, and multiple pulses per period. For \MASS\ we determine a pulse period consistent with previously determined values.  We modeled locations of pulsed emission using an oblique rotating magnetospheric model with beamed electron cyclotron maser (ECM) sources. The best-fit models have narrow ECM beaming angles aligned with the local source magnetic field direction, except for one isolated burst from \MASS. For \TVLM, the best-fit rotation axis inclination is nearly orthogonal to the line of sight.  For \MASS\ we found a good fit using a fixed inclination $i=36\degr$, determined from optical observations. For both stars the ECM sources are located near feet of magnetic loops with radial extents $1.2R_s-2.7 R_s$ and surface fields 2.2 - 2.5 kG. These results support recent suggestions that radio over-luminous UCDs have a global `weak field' non-axisymmetric magnetic topologies.
\end{abstract}
\section{Introduction} 
The discovery of intense, non-thermal radio emission from stars at the low-mass end of the main sequence \citep[e.g.,][]{Berger:2001, Berger:2002, Hallinan:2006, Osten:2006a, Phan-Bao:2007}  implies the presence of strong magnetic fields.  These fields are unexpected given the fully convective stellar interior and observed sharp decline in chromospheric H$\alpha$, and coronal X-ray emission for dwarf stars \citep[e.g.][]{Neuhauser:1999, Gizis:2000, West:2004}. The $\alpha$-$\Omega$ dynamo, driven by shearing motions at the radiative-convective boundary, is the canonical model for magnetic field generation in solar-type stars, but this mechanism clearly cannot apply to fully convective dwarfs. Instead, magnetic field in these stars may be generated by $\alpha^2$ or turbulent dynamos, which are driven by turbulent motions associated with internal convection or on both stratification and rotation \citep{Raedler:1990, Durney:1993, Browning:2008}. 

Radio surveys of ultra-cool dwarfs (UCDÕs, spectral class M8 and cooler) have found that about 10\% of these system are radio luminous \citep{Berger:2006, Antonova:2008, Antonova:2013}. The radio luminosity of the detected systems is far in excess of the well-known G\"udel-Benz  relation \citep[][GB]{Gudel:1993a}, an empirically-derived ratio between radio and X-ray luminosity (log Lr/Lx $\sim$ -15.5)  that applies to magnetically active stars over a wide range of spectral types. A theoretical model that has been suggested to explain the GB correlation is the chromospheric evaporation model \citep{Machado:1980,Allred:2006}, in which x-ray emission results from heating and evaporation of chromospheric plasma caused by non-thermal beamed electrons, which produce gyrosynchrotron radio emission \citep{Neupert:1968}. However, the dozen UCDs detected in the radio to date \citep{Berger:2002, Berger:2006, Burgasser:2005, Phan-Bao:2007,Antonova:2008, McLean:2011, Mclean:2012, Route:2012} all violate the G\"udel - Benz relation by orders of magnitude, suggesting that the chromospheric evaporation model does not apply to these stars. 
	
In 2006, an even more unexpected discovery was made: A few UCD stars had periodic, pulsar-like radio emission \citep[e.g.,][]{Hallinan:2007, Berger:2009}. The pulsed emission was 100\% circularly polarized and occurred either once or twice per rotational period, depending on the observed frequency. The emission mechanism was attributed to the electron-cyclotron maser instability \citep[ECM,][]{Hallinan:2008}, which would account for both the high circular polarization and the apparent beaming of the radiation. 

The ECM mechanism has been well-studied in space and planetary environments, both theoretically \citep[e.g.,][and references]{Treumann:2006} and observationally \citep[e.g.,][]{Zarka:2004}. The ECM instability couples the kinetic energy of electrons spiraling in converging magnetic fields to the ambient radiation field at the electronÕs gyro frequency, resulting in exponential growth of the radiation field at the local gyro-frequency. The radiation pattern is strongly beamed, since the growth rate is strongly peaked for wave vectors oriented normal the the magnetic field \citep{Mutel:2007}. However, ECM growth is quenched unless the ratio of the electron plasma frequency is much less than the cyclotron frequency  Ñ this implies that the source region has a high magnetic field and/or low density plasma. 

In a planetary magnetosphere such as the Earth, the ECM growth rate is highest in density-depleted auroral cavities at high magnetic latitudes \citep{Ergun:1998, Ergun:2000}, resulting in radiation at frequencies between 50 ~KHz and ~500 KHz, i.e. magnetic fields between 0.02 and 0.2 Gauss. The radiation is dominantly in the extraordinary mode and is initially linearly polarized (Ergun et al. 1998), but becomes circular as the radiation is refracted upward and parallel to the magnetic field.  The beaming pattern of terrestrial ECM emission, known as auroral kilometric radiation, is strongly modified by the auroral cavity, so that the resulting far-field pattern resembles a cigar-shape, with the long axis parallel to the auroral cavity \citep{Mutel:2008}. ECM emission at both Jupiter \citep[e.g.][]{Imai:2008} and Saturn \citep{Lamy:2011} is also highly beamed, although in these planets it is not yet established whether density cavities are responsible for the beaming. Whether this also occurs in stellar magnetospheres is not known.

For UCD stars, the pulses have been observed between 1.4~GHz and 10~GHz, which implies magnetic fields of order several kilogauss. These strong magnetic fields, although surprising in fully convective stars, may not be completely unexpected. \citet{Reiners:2007} studied the magnetically sensitive Wing-Ford FeH band in a wide spectral range of M dwarfs, and found kilogauss surface fields, thus extending the direct measurement of magnetic field strengths to spectral type M9. 

\citet{Morin:2010} proposed that late-M dwarfs have magnetic field topologies that can be classified into one of two types: (1) strong axisymmetric dipolar fields or (2) weak non-axisymmetric fields, with perhaps strong localized regions. It has been suggested that these two topologies can be associated with distinctly different emissions. The  UCDs that are radio-dim, X-ray bright, and closely follow the GB relation are believed to have strong axisymmetric fields, while the radio-bright, X-ray dim UCDs that break the GB relation are thought to have weak non-axisymmetric fields \citep{Mclean:2012, Williams:2014}. Furthermore, recent numerical models of M-dwarf dynamos find that bistability of the magnetic field is common within the parameter range covered in the models. This supports the notion that both a dipole-dominated and multipolar field are possible configurations for stars with similar physical parameters \citep{Morin:2011, Gastine:2013}.

This paper reports multi-epoch radio observations of two well studied UCDs, 2MASS J0746425 +200032 (hereafter \MASS) and TVLM 513-46546 (hereafter \TVLM). We used the new wideband capabilities of the Karl G. Jansky Very Large Array\footnote{The National Radio Astronomy Observatory is a facility of the National Science Foundation operated under cooperative agreement by Associated Universities, Inc.} (VLA) to characterize both the temporal and spectral properties of the continuum and pulsed radio emission.  For both stars, we constructed  time-frequency (a.k.a. dynamic) spectra of Stokes I and V emission over several pulse periods at multiple epochs.  We compared the pulse morphologies in these spectra  with synthetic dynamic spectra generated using an oblique rotating magnetospheric model. By fitting the geometrical and beaming free parameters of the model, we determined possible source locations for the pulsed emission regions. We also fit spectral energy distributions of the non-pulsed continuum emissions to power-law gyrosynchrotron models  to estimate the number density and mean magnetic field strength in the magnetosphere.

\subsection{Target Star Properties}
\MASS\ is a  dwarf binary system (L0 + L1.5) at a distance of 12.2 pc \citep{Dahn:2002}. The orbit is elliptical \citep[$e=0.49$,][]{Konopacky:2010} with a semi-major axis  2.7~AU \citep{Reid:2001}. The  orbital inclination of the binary system is 41.8\degr $\pm$ 0.5\degr  \citep{Konopacky:2012}, and the equatorial and orbital planes are coplanar within 10\degr \citep{Harding:2013}. In addition to continuous radio emission \citep{Antonova:2008},  \MASS\ has pulsed radio emission with a 2.07 hour  period \citep{Berger:2009}. \citet{Harding:2013} recently analyzed \vsini and periodic light curve variations and determined that the primary has a 3.32 hour period, implying that the less-massive secondary must be the origin of the 2.07 hr periodic radio emission.  

\TVLM\ is an M9 dwarf located at a distance of 10.6 pc \citep{Dahn:2002}. It was first detected at radio wavelengths by \citep{Berger:2002} and has been extensively studied since \citep[e.g.,][]{Osten:2006b,Hallinan:2006,Berger:2008a}. \citet{Hallinan:2007} found periodic, highly polarized radio pulses and both 5~GHz and 8~GHz, but with opposite circular polarizations. \citet{Berger:2008a} confirmed the presence of polarized flares at 8~GHz. and showed the pulse period was the same as periodic changes in $H_{\alpha}$ equivalent width, although intensity variations of individual pulses  were not correlated with $H_{\alpha}$ variations. They suggested that both the pulsed radio and optical emission originates in a co-rotating chromospheric hot spot or an extended magnetic structure with a covering fraction $\sim50$\%. 

\section{Observations and Data Analysis}

\begin{deluxetable*}{lccccc}
\tablewidth{4.5in}
\tablecolumns{6}
\tablecaption{ VLA Observing Log} 
\tablehead{ \colhead{Source} & \colhead{Epoch} & \colhead{Array} & \colhead{UT Range} & \colhead{Lower band} & \colhead{Upper band} \\
 & & & & GHz & GHz } 
\startdata

\MASS & 2010 Nov 12 & C & 08:07 -- 12:36 & 4.2 -- 5.2 & 6.8 -- 7.8 \\ 
\MASS & 2010 Dec 16 & C & 04:20 -- 11:24 & 4.2 -- 5.2 & 6.8 -- 7.8 \\ 
\MASS & 2012 Dec 08 & A & 06:24 -- 08:58 & 4.4 -- 5.4 & 6.4 -- 7.4 \\ 
\MASS & 2012 Dec 11 & A & 06:26 -- 08:02 & 4.4 -- 5.4 & 6.4 -- 7.4 \\ 
\MASS & 2012 Dec 14 & A & 05:30 -- 08:05 & 4.4 -- 5.4 & 6.4 -- 7.4 \\ 

\TVLM & 2011 May 07 & B & 06:25 -- 12:38 & 4.2 -- 5.2 & 6.8 -- 7.8 \\ 
\TVLM & 2011 May 08 & B & 06:21 -- 12:34 & 4.2 -- 5.2 & 6.8 -- 7.8 \\ 
\TVLM & 2011 July 02  & A & 02:45 -- 08:58 & 4.2 -- 5.2 & 6.8 -- 7.8 \\ 
\TVLM & 2012 Dec 22 & A & 02:52 -- 15:32 & 4.4 -- 5.4 & 6.4 -- 7.4 
\enddata
\label{table:vla-obs}
\end{deluxetable*}

We used the wideband WIDAR correlator system at the VLA to observe \MASS\ at three epochs (8, 11, 14 Dec 2012) and \TVLM\ at one epoch (22 Dec 2012), all during A array configuration. We observed simultaneously in two 1~GHz wide bands centered on 5.0 and 7.0 GHz. The receiver bandpass correction and absolute flux density was set using the amplitude calibrator 3C48. The angularly nearby source J0738+1472 was used for phase calibration. In addition, we calibrated and analyzed five archival wideband VLA observations of these stars (10B-209, Table\ref{table:vla-obs}) . These observations also used two 1~GHz bands, centered on 4.8 and 7.5 GHz. The data editing, calibration, and mapping was done using NRAO's  Common Astronomy Software Application (CASA) using standard data reduction techniques. Observing details are summarized in Table~\ref{table:vla-obs}.

After editing and calibrating, we made Stokes I and V `snapshot' maps for each source and epoch using time-frequency sampling intervals of 1 minute and 128~MHz. We then 
determined the peak flux density in a box centered on the source position and with dimensions equal to the restoring beam (CASA task IMSTAT).  These binned data samples were used to construct time-frequency plots for both sources. 

As discussed above, the radio emission has both a continuum and a pulsed component. In order to visually accentuate the (more intense) pulsed component in dynamic spectra plots, we filtered the binned data, replacing low signal-to-noise (SNR) bins (I $< 6\sigma$ , V/I $<4\sigma$) with zero flux density.  On the other hand, In order to determine the spectral characteristics of the continuum emission, we analyzed only those time intervals with no pulsed emission. 

\section{RESULTS}

\subsection{Dynamic spectra}

For both \TVLM\ and \MASS, the dynamic spectra reveal pulsed emission with a complex, epoch-dependent frequency dependence. The temporal variations are presumably caused by the emergence and decay of active regions in each star's magnetosphere that are the origin of the pulses. The characteristic lifetime of these regions, inferred from the persistence of the pulse profiles, is of order months to several years. In section~\ref{sec:pulse-model} we model the the location and physical environment of these active regions.

\begin{figure*}
\epsscale{1.19}
\hspace*{-0.05in}
\plotone{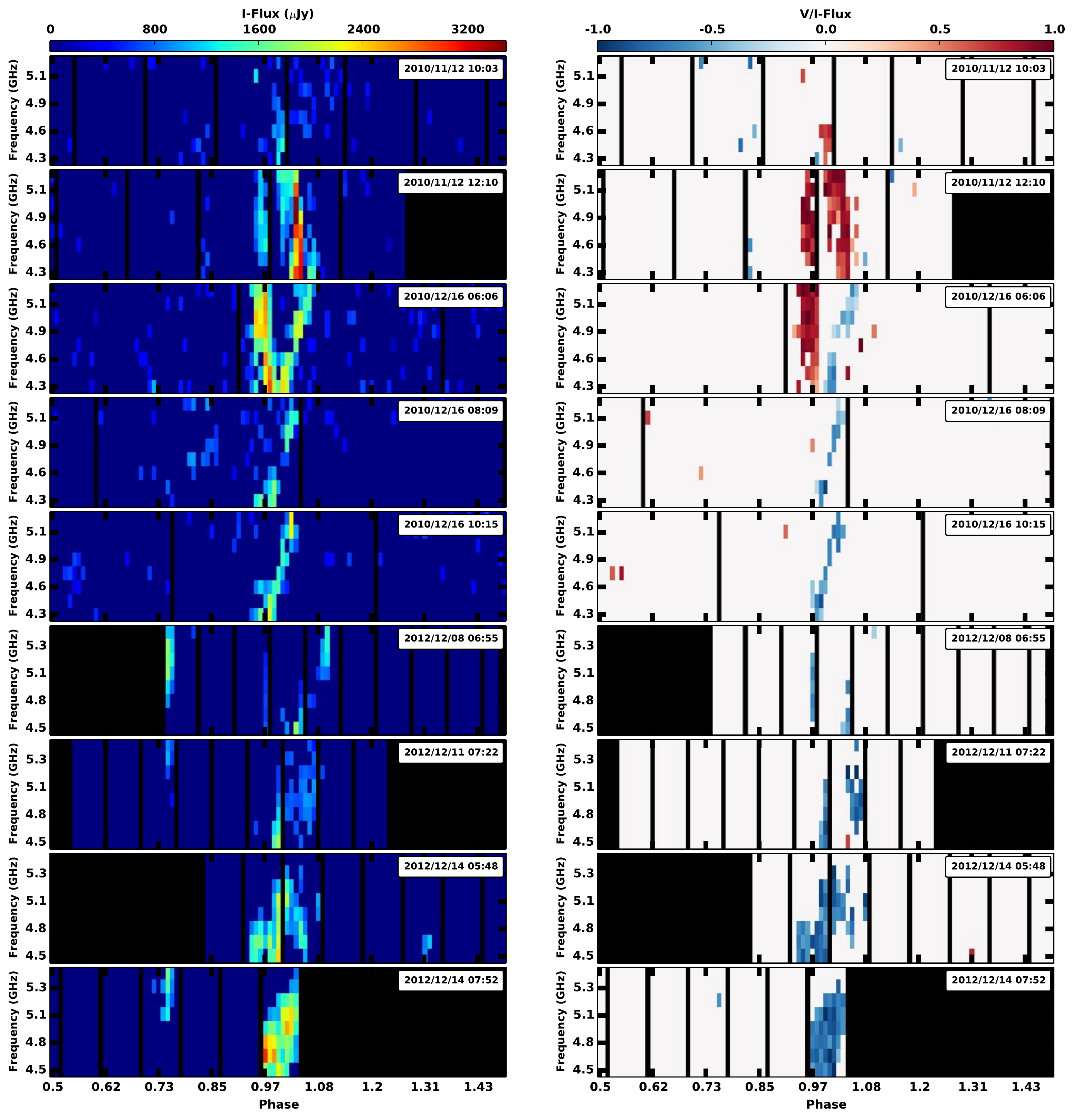}
\caption{Composite dynamic spectra of \MASS\ for each of the observations; the date of the observation and UT of the main pulse are given in the inset.  The left panel displays Stokes I spectra while the right panel shows Stokes V/I, with blue indicating The phases are calculated using the period derived in section~\ref{sec:period}.  }
\label{fig:2M0746-dynspectra}
\end{figure*}

Dynamic spectra of the star \MASS\ in Stokes I and V/I are shown in Fig.~\ref{fig:2M0746-dynspectra} for five epochs from 12 Nov 2010 to 14 Dec 2012. Note that for several epochs more than one pulse is shown. These correspond to observing times spanning several pulses periods. Also, only the lower observed frequency band is shown, since there was no detectable radio emission in the upper band except at epoch 2010 Dec 16, which we discuss separately in section~\ref{sec:pulse-model}. 

The spectra have been phase-shifted so that the brightest pulse in each spectrum is located at phase 0 on average. These pulses which we refer to as the main pulse, show a clear evolution with epoch. The pulse phase was computed using the period 2.071478 hr (see section~\ref{sec:period}). At early epochs, the main pulse is weak and right-circularly polarized, but becomes left-circularly polarized with a frequency drift $df/dt\sim2$ MHz/sec starting at epoch 2010 Dec 16. This frequency drift becomes a stable feature of the main pulse for several years, up through at least the last observing epoch on 14 Dec 2012. There is also a secondary pulse at phase $\sim0.75$ at later epochs (2012 Dec 8,11,14).  Curiously, this pulse is not significantly polarized. This is similar to the UCD star 2M 0036+182, which \citet{Hallinan:2008} found has a highly circularly polarized main pulse but a largely unpolarized interpulse.
 
\begin{figure*}
\centering
\epsscale{1.19}
\hspace*{-0.05in}
\plotone{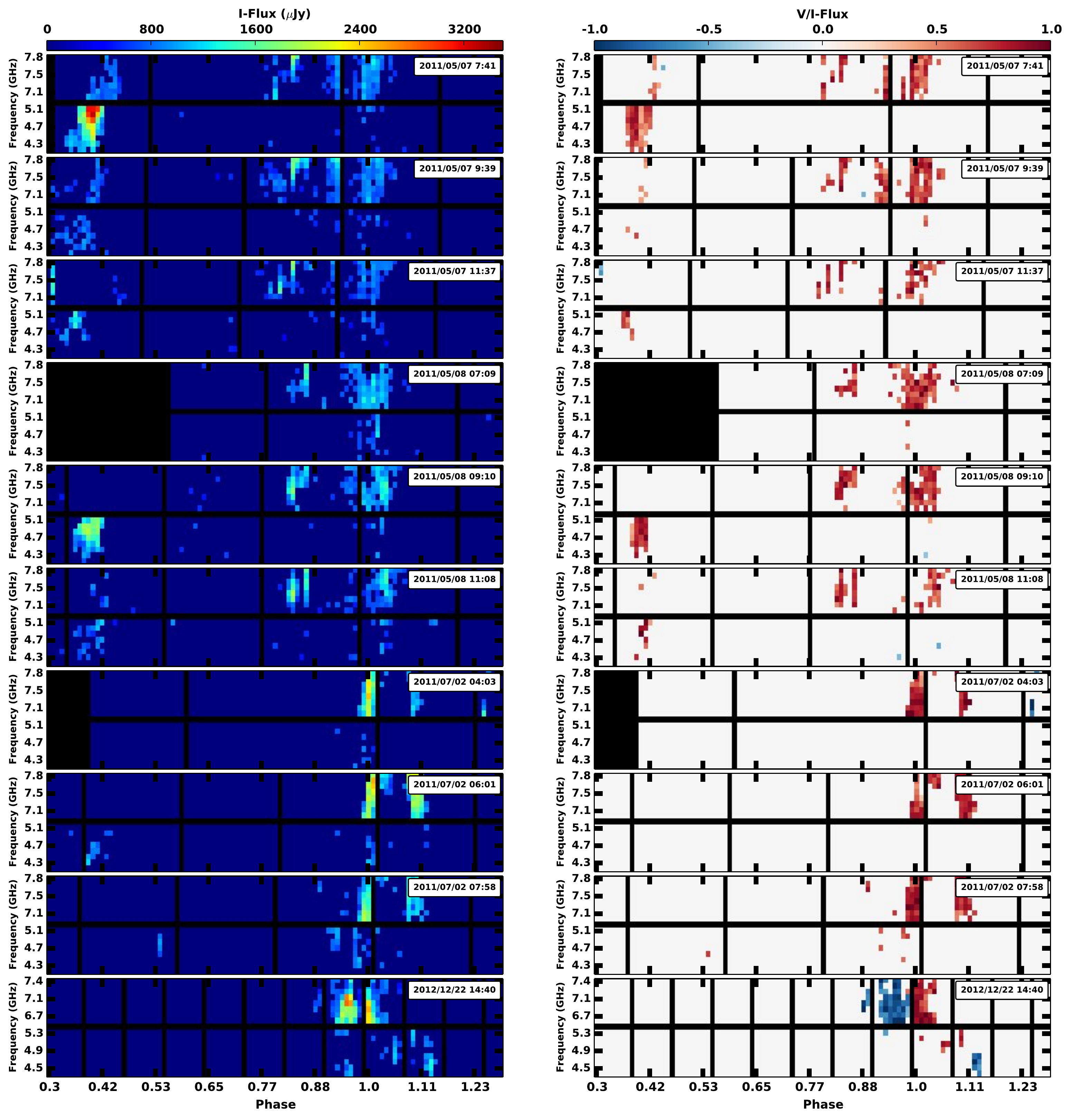}
\caption{Composite dynamic spectra of \TVLM\ for each of the observations; the date of the observation and the UT of the MP is given in the inset. The left panel show the Stokes I spectra while the right panel shows Stokes V/I. The phases are calculated using the period derived in section~ref{sec:period}.  }
\label{fig:TVLM513-dynspectra}
\end{figure*}

\TVLM\ dynamic spectra for four epochs is shown if Fig.~\ref{fig:TVLM513-dynspectra}, with several epochs with longer observing intervals showing multiple pulses. As with \MASS, we phase-shifted the strongest pulse using a calculated period 1.95950 hr, as described in section~\ref{sec:period}. Both the low and high bands are shown for \TVLM\ since there is detectable emission in both bands at all epochs. The main pulse, which is nearly always right-circularly polarized, is brightest in the upper band, but is also detectable in the lower band. It is also almost always a double pulse, with a phase separation $\sim0.1$ phase. There is also a secondary pulse near phase 0.35 seen most intensely in the lower band at three epochs . 

These pulse characteristics are significantly different from previously published pulse profiles for \TVLM. For example, \citet{Hallinan:2007} detected two repeating LCP pulses at 4.9~GHz, but separated by $\sim0.5$ in phase, and two oppositely polarized pulses at 8.4~GHz, separated by $\sim0.1$ phase. The pulse characteristics are also systematically different from \MASS\ in that we detect the strongest pulses at much higher frequencies and do not detect any frequency drift. We discuss implications of these differences in terms on different rotation axis inclinations in section~\ref{sec:pulse-model}

\subsection{Pulse Periods}\label{sec:period}
In this section, we combine our observed main pulse (MP) times with previously published MP timings to calculate an improved estimate of the pulse period. To measure the observed MP times we created sets of CLEAN maps with 1~GHz spectral and 10 second time averaging for both the upper and lower bands of each of the observations. We then fit the source in each map using IMFIT and created light curves for each observation in both spectral bands.  The MP peaks in each light curve were then fit with a gaussian to give time of the central peak and the width of the peak, which is taken as the uncertainty in the MP time. The uncertainties in the MP time ranged between 2-3 minutes depending on the observation epoch. Heliocentric corrections were applied to the MP times before proceeding with the period estimate.

To estimate the pulse period, we compute the sum of the absolute value of the phase difference between the observed and model MP weighted by the reciprocal timing uncertainty ($\sigma$) for each observed MP, 
\begin{equation}
f(\epsilon)=\displaystyle\sum_{1}^{N}
\frac
{|\rm{min}[\phi_i(\epsilon), 1- \phi_i(\epsilon)]|}
{\sigma_i},
\end{equation}
where min[] is the minimum function and $\phi_i(\epsilon)$ is the observed MP phase at epoch $i$ computed using a trial period $P(\epsilon) = P_0 + \epsilon$, where $P_0$ is the previously published period for each star. We varied  $\epsilon$ over the range $\pm10$ sec in increments of 1~msec to find a minimum in the difference function $f(\epsilon)$. 

It is important to note that for both stars, the observed MP timing datasets are unevenly sampled with several large gaps of several thousand periods between sampled epochs. For such datasets, it is difficult to determine whether phase jumps occurred in the gaps. This is a result of the ambiguity concerning the number of pulse counts between observations. Consider a series of consecutive pulses separated by a nominal period $P$ and uncertainty $\sigma$. Now consider another series of pulses with the same period and uncertainty, but observed $n$ periods later, where $n\gg1$. The latter pulses can be rectified with the previous series by fractionally adjusting the period by $1/n$. If $\sigma>1/n$, all pulses will be compatible with the adjusted period {\it whether or not there was a phase jump in the gap}. Hence, the period estimates given below may not provide a reliable ephemeris for prediction of the MP arrival time at future epochs. 

\begin{figure}
\centering
\epsscale{1.2}
\hspace*{-0.1in}
\plotone{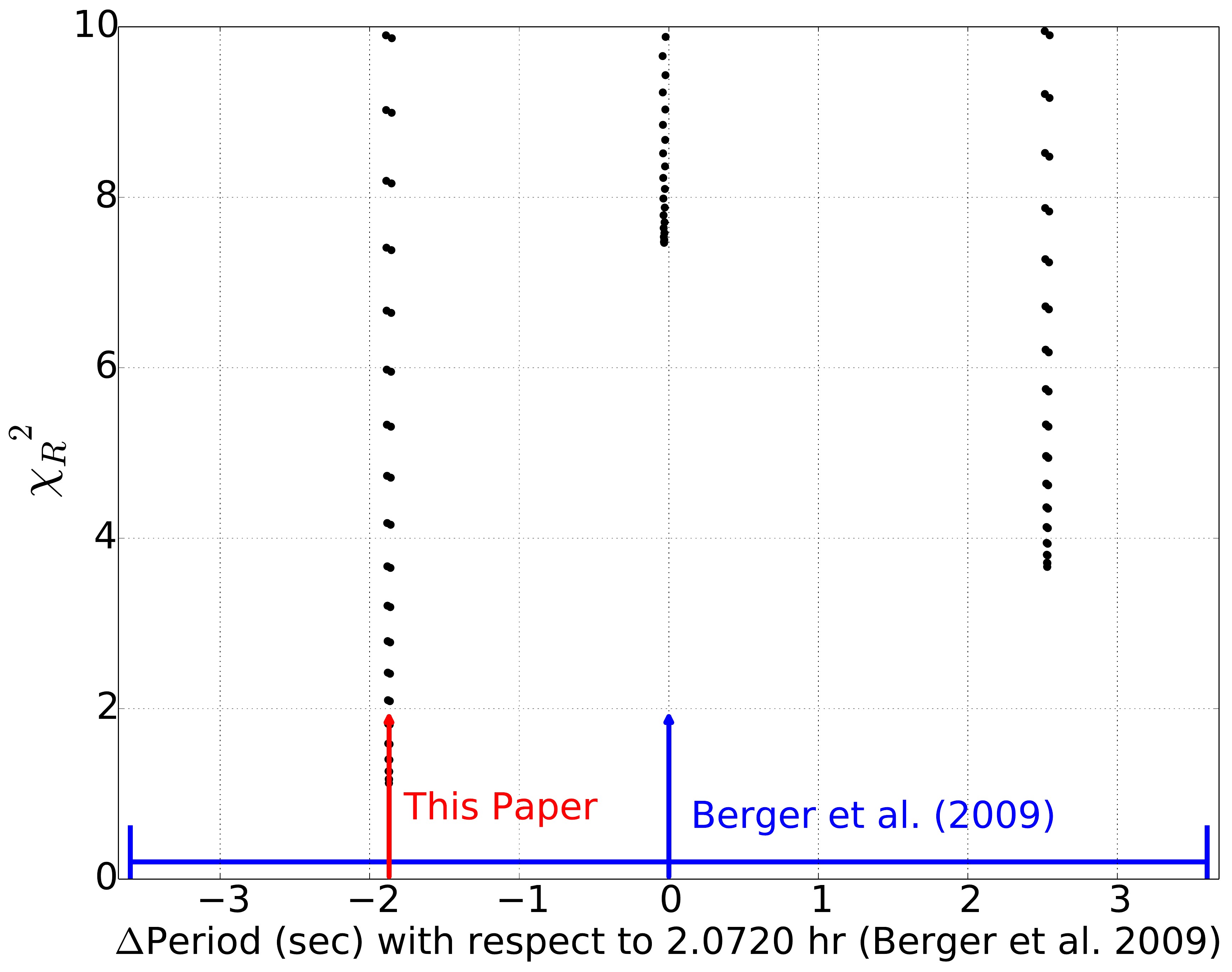}
\caption{Normalized chi-square of the summed difference between the published main pulse period  \citep[P=2.072 hr,][]{Berger:2009} and trial period for \MASS. The lowest minimum corresponds to a correction $\epsilon=-1.867$ seconds giving a revised period 2.071480 hr $\pm$ 0.000002 hr. This value lies within the period uncertainty range of \citet[][blue bar]{Berger:2009}.}
\label{fig:2MPhDiff}
\end{figure}

\subsubsection{\MASS\ pulse period}

From the observations presented here for \MASS\ we use only the MP times for the Dec 2010 and Dec 2012 observations since these pulses all have the same structure and helicity of its circular polarization. In addition to the MP times determined in this study (shown in Fig.\ref{fig:2M0746-dynspectra} insets), we used MP times from \citet{Berger:2009}. We note that although \citet{Antonova:2008} also observed a radio pulse from \MASS, we did not use it because it was unclear whether the pulse was a MP or an inter pulse. 
We used the published period of \citet{Berger:2009} 2.072$\pm$0.001 hr as a starting point for the minimization search. The summed differences as a function of $\epsilon$ are shown in Figure~\ref{fig:2MPhDiff}. The minimum value corresponds to a correction $\epsilon =-1.876 $ seconds,  resulting in a revised period 2.071480 $\pm$ 0.000002 hr where the error is determined from a $\chi^2$ analysis of the residuals. This period is consistent with the period uncertainty determined by \citet{Berger:2009} but improves its precision. 

\begin{figure}
\epsscale{1.19}
\centering
\plotone{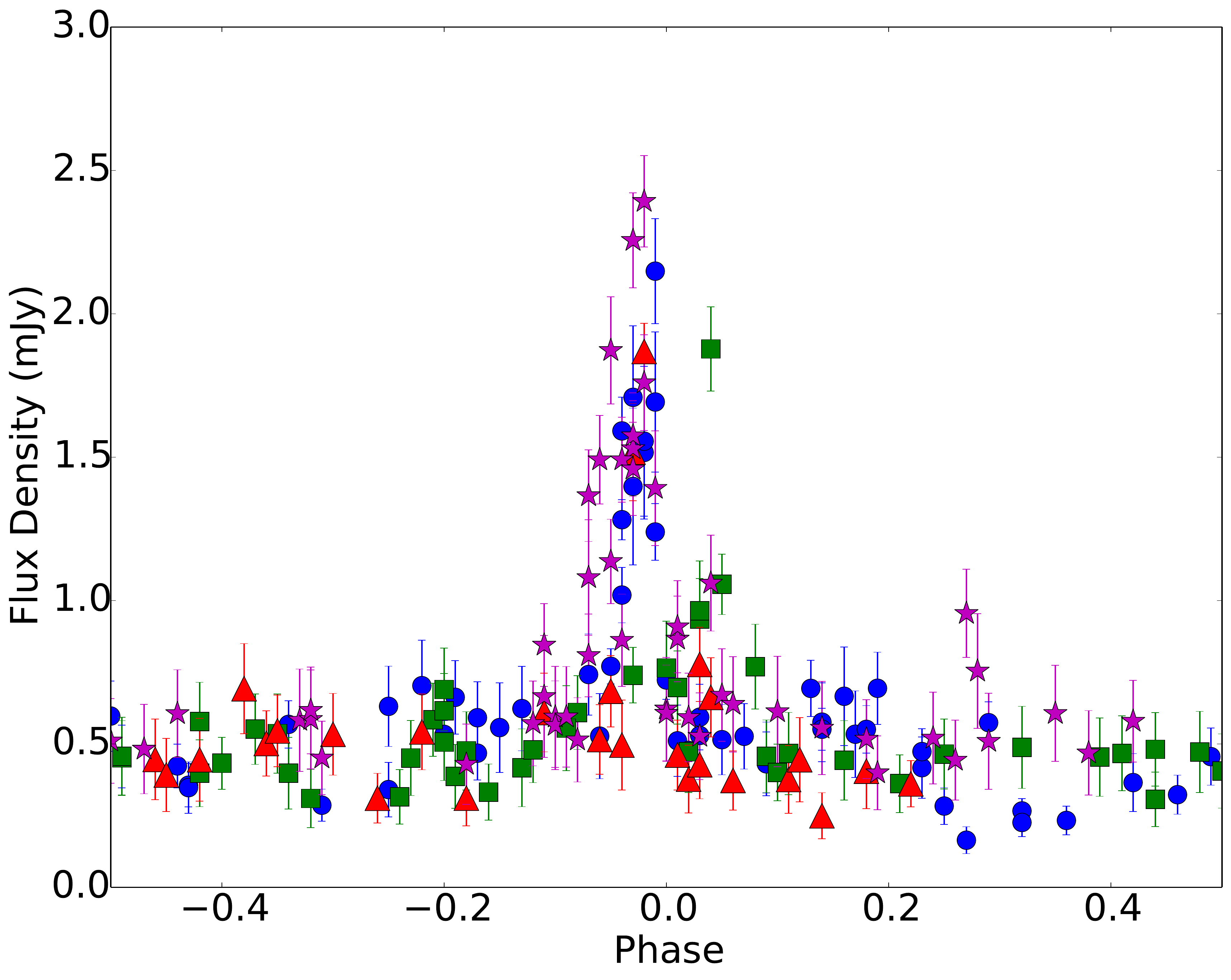}
\caption{ \MASS\ phase folded light curves  for epochs 2010 Dec 16  (blue circles), 2012 Dec 08 (green squares), 2012 Dec 11 (red triangles), and 2012 Dec 14 (magenta stars), computed using the best fit period 2.07148 hr.  Each point is an averaged 60~s x 128~MHz time-frequency  Stokes I flux density, with a central frequency of 4.5 GHz  (2012 epochs) and 4.3 GHz   (2010 epochs). }
\label{fig:2MPhaseFold}
\end{figure}

To check the period stability for \MASS, we phase-folded the observed light curves using the best fit period  (Figure \ref{fig:2MPhaseFold}). While the 2010 Dec 16,  2012 Dec 11, and 2012 Dec 14 pulses are aligned in phase, the 2012 Dec 08 pulse is slightly shifted in phase. This small phase difference could be due to a small change in the longitude of the active region or from a small beaming angle change. However, we think it is unlikely that a longitudinal shift occurred since the observed  pulses in the latter two observations (2012 Dec 11 and 14) both align with the pulses of the 2010 observation.

\subsubsection{\TVLM\ pulse period}

For \TVLM\ there are several previously published rotational periods from both radio and optical observations \citep{Hallinan:2007, Berger:2008a, Doyle:2010, Harding:2013a, Wolszczan:2014}.  The most recent period analysis by \citet{Wolszczan:2014} and \citet{Harding:2013a} find nearly identical pulse periods that are both 30~s shorter than the previous  estimate of \citet{Doyle:2010}. \citet{Wolszczan:2014} suggest that this discrepancy is most likely the result of a significant phase shift between 2007 April and 2007 June, which is unaccounted for in the \citet{Doyle:2010} analysis. 

\begin{figure}
\epsscale{1.19}
\centering
\plotone{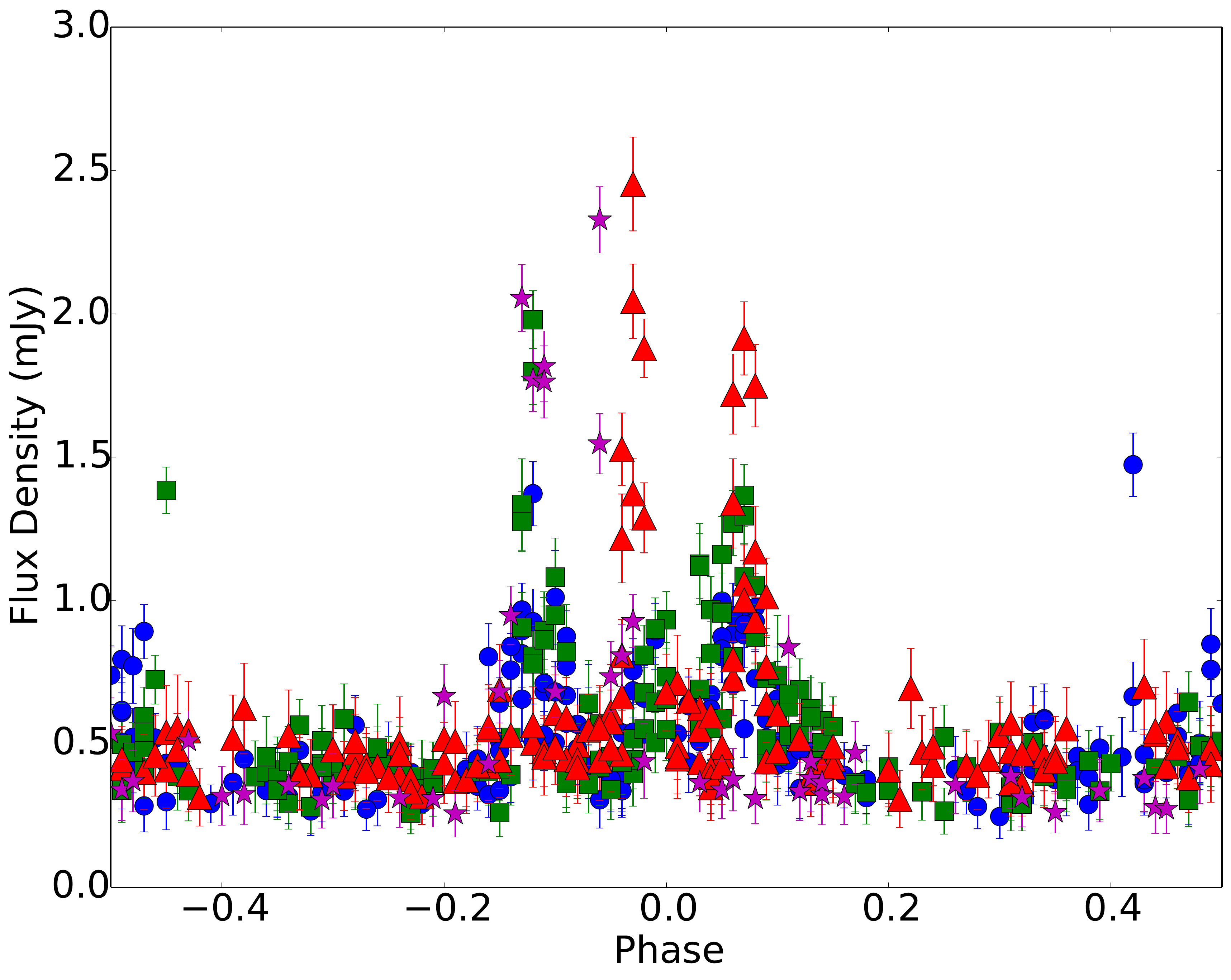}
\caption{ \TVLM\  phase folded light curves for epochs 2011 May 07 (blue circles), 2011 May 08 (green squares), 2011 July 02, and 2012 Dec 22 (magenta stars)  using the best fit period of \citet{Wolszczan:2014}.  Each point is an averaged 60~s x 128~MHz time-frequency Stokes I flux density, with a central frequency  6.6 GHz  (2012 epochs) and 7.3 GHz  (2011 epochs).}
\label{fig:TVLMPhaseFold}
\end{figure}

\citet{Wolszczan:2014} find a best fit period 1.959574$\pm$0.0000002~hr  using pulse timings over a 7~yr time span, indicating a stable period over this time range. Figure \ref{fig:TVLMPhaseFold} shows  phase-folded  light curves for the  \TVLM\  epochs reported in this paper using the \citet{Wolszczan:2014} period. The pulse profile is complex and epoch-dependent but  is phase-stable, confirming the constancy of the \citet{Wolszczan:2014} period during epochs 2011 and 2012.

\citet{Wolszczan:2014} also report that their radio period is the same as the optical period determined by \citet{Harding:2013} but with a phase offset $0.41\pm.02$ (radio preceding optical).  By coincidence, our 2011 May 7,8 observations were coincident with those of \citet{Harding:2013}. We visually estimated the  times of brightness maxima  from their published light curves. We also find a phase offset  0.4$\pm$0.2 phase, consistent with the radio-optical offset reported by \citet{Wolszczan:2014}.

\subsection{Quiescent emission}\label{sec:quiescent-obs}
In addition to pulsed emission, broad-band quiescent radio emission is detected from each source. To study this component, we first identified time intervals that do not include the bright pulsed emission (I $>6\sigma$) in any of the 128~MHz frequency bins. Cleaned maps were made by averaging over these  time intervals in 512~MHz frequency windows.  In each frequency window we measured the peak Stokes I and V  flux density by fitting Gaussian profiles at the source location using the CASA task IMSTAT.  Figure~\ref{fig:GSModel} shows the continuum Stokes I  spectral energy distributions for each source at all observed epochs (cf. Table~\ref{table:vla-obs}), along with gyro-synchrotron model fits as described in section~\ref{sec:gs-model} below.

\begin{deluxetable}{ll}[t!]
\tablewidth{3.0in}
\tablecolumns{2}
\tablecaption{ Quiescent emission spectral Indices}
\tablehead
{ \colhead{Epoch}& \colhead{$\alpha_{4.5-7.5\ \text{GHz}}$}  }
\startdata
\cutinhead{\MASS}
2010 Nov 12&-0.42 $\pm$ 0.09 \\ 
2010 Dec 16&-0.34 $\pm$ 0.09\\ 
2012 Dec 08&-0.34 $\pm$ 0.30\\ 
2012 Dec 11&-0.72 $\pm$ 0.30\\ 
2012 Dec 14&-0.53 $\pm$ 0.20\\
\cutinhead{\TVLM}
2011 May 07&-0.45 $\pm$ 0.06\\ 
2011 May 08&-0.35 $\pm$ 0.06\\ 
2011 Jul 02&-0.44 $\pm$ 0.08\\ 
2012 Dec 22&-0.22 $\pm$ 0.10
\enddata
\label{table:QE}
\end{deluxetable}
\subsubsection{Spectral indices}
We used the peak I-flux measurements in the four 512~MHz bins to determine the spectral index for each observation (Table~\ref{table:QE}).  The spectral indices  are consistent with previously published measurements of -0.4 $\pm$ 0.1 \citep{Osten:2006b} for \TVLM\ and -0.7 $\pm$ 0.3 \citep{Berger:2009} for \MASS. 
\subsubsection{Circular polarization}\label{sec:GS-Pol}
Circular polarization (Stokes V/I, hereafter $\pi_c$) was not detected from \TVLM\ for any of the quiescent emission observations, to a limiting fraction between 0.1 and 0.2, depending on epoch,  This is in accord with \citet{Osten:2006} who found $\pi_c < 0.15$ at both 5~GHz and 8.4~GHz. Likewise, we did not detect significant polarization from \MASS, except at epochs  2010 Nov 12 and 2010 Dec 16, where we detected  $\pi_c \sim 0.20$  in the lower frequency band and place upper limits of $\pi_c\sim0.13$ at higher frequencies. However, we believe this emission to be contaminated by low level pulses and take the measurements to be upper limits on the polarized emission. Assuming gyrosynchrotron emission, this provides an approximate estimate of the angle-averaged energetic electron energy, 
\begin{equation}
\langle E \rangle \sim \frac{1}{\pi_c} m_ec^2 \sim \rm{2.5~MeV.}
\end{equation}

\begin{figure*}
\begin{center}
\includegraphics[width=6.5in]{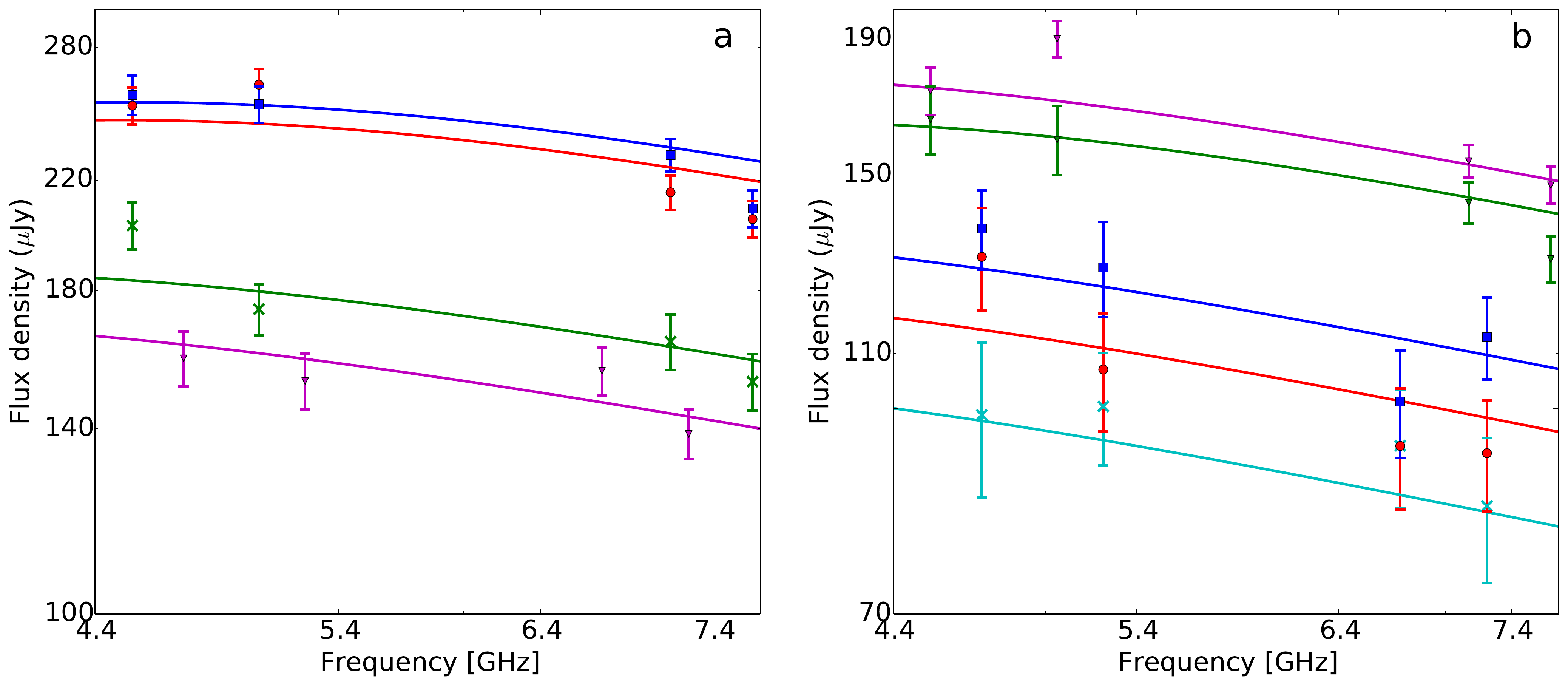}
\end{center}
\caption{Comparison between model and measured quiescent emission spectral energy distributions (SED). (a) \TVLM\ peak Stoke I flux density for: May 7 2011 (blue), May 8 2011 (red), July 2 2011 (green), and December 22 2012 (magenta). The model SED's are shown by  solid lines, where line are color-coded  by epoch. (b) Same as (a), but for \MASS\ where epochs are  November 12 2010 (green), December 16 2010 (magenta), December 8 2012 (cyan), December 11 2012 (red) and December 14 2012 (blue).}
\label{fig:GSModel}
\end{figure*}

\subsubsection{Brightness temperature}
The brightness temperature of an incoherent radio source as a function of distance $d$, flux density $S_{\nu}$, frequency $\nu$ , and effective diameter $D$ can be conveniently expressed as,
\begin{equation}
T_b = 10^{10}\left(\frac{S_{\nu}}{\text{mJy}}\right)\left(\frac{\nu}{\text{GHz}}\right)^{-2}\left(\frac{d}{\text{pc}}\right)^{2}\left(\frac{D}{R_{\text{J}}}\right)^{-2}\ \text{K},
\end{equation}
 where $R_{J}$ is the radius of Jupiter. Using the measured quiescent emission flux densities and assuming  the emission region sizes are of order a stellar diameter,  the brightness temperatures for both UCDs  are in the range $(1-5)\times\ 10^9$ K. 
 
 These high brightness temperatures, combined with the  fractional circular polarization detected in \MASS\ and  the power-law spectra all rule out thermal emission, but support a model of optically-thin gyro-synchrotron radiation from a population of mildly relativistic power-law electrons (see section~\ref{sec:gs-model}).
 
We note that  the spectral energy distributions for both stars are relatively stable over years. This long-term stability has also been observed for quiescent emission form other UCDs \citep[e.g.,][]{Osten:2009}, and is in marked contrast to other radio-loud stellar sources, such as RS CVn binaries. A power-law gyrosynchrotron process is also usually invoked for these sources, but unlike the UCD's, they produce large flares, with dramatic changes in  flux density, spectral index, and fractional polarization \citep[e.g.,][]{Mutel:1998, Richards:2003}. This may reflect a difference in energization, wherein flares in active binaries may be energized by large-scale magnetic interactions between components \citep{Richards:2012}, while  UCD's may be energized by  quasi-continuous, low-level magnetic reconnection events \cite{Williams:2014}.

\section{Coronal Models}
\subsection{Pulsed Emission: ECM sources on isolated loops}\label{sec:pulse-model}
The temporal behavior of frequency and polarization of ECM-driven pulses provide robust constraints on the  topology of the stellar magnetosphere, provided the source-dependent parameters (e.g. angular beaming, refraction) can be tenably modeled. The key idea is that each ECM emission frequency  directly maps the local magnetic field strength, and the rotation of the star provides time-lapsed spatial slices of the region of the magnetosphere favorably aligned to the beamed radiation. 

The properties of ECM-driven radiation depend on the plasma conditions at the emission site. For low density plasma (ratio of electron plasma to cyclotron frequency  $\omega_{pe}/\Omega_{ce}<<1$), and a loss-cone electron phase distribution, the dominant mode is R-X at the fundamental harmonic. This implies the radiation is right-circularly polarized  at a frequency very close to the electron cyclotron frequency. This is the most commonly observed ECM mode in planetary magnetospheres \citep[e.g.][]{Zarka:1998}. However,  for stellar magnetospheres, this may not apply. For example, stellar ECM regions could have larger plasma to cyclotron frequency ratios that result in higher harmonics and/or L-O mode being the dominant emission \citep{Lee:2013}. Since we have no independent information on the plasma conditions at the ECM sites, we have applied Occam's razor and assumed the planetary case for the model viz. R-X mode at the fundamental harmonic. 

We modeled the observed dynamic spectra with a set of ECM emission sources fixed in an oblique rotating coordinate system (i.e., the star's rotation axis inclination). Since the pulses have a short duty cycle, we assume there exists a small number of localized magnetic loops on which the ECM instability is active. Each source location is chosen so that the emission frequency is equal to the local electron gyro-frequency. Since unstable electron beams will populate an entire L-shell field line, we also assume that there exists a pair of ECM sources at conjugate points on each active L-shell field line. As viewed by an observer whose location intercepts the angular pattern of both sources, the pulses from conjugate locations will have opposite circular polarizations. For globally dipolar fields,  helicity is used to map opposite magnetic hemispheres e.g. at Saturn \citep{Lamy:2008}.

The angular beaming pattern of stellar ECM emission is unknown, but multi-spacecraft studies of terrestrial ECM emission \citep[known as auroral kilometric radiation,][]{Gurnett:1974} have shown that the emission is elliptically beamed with the major axis zonally aligned, i..e, along the density-depleted auroral cavities \citep{Mutel:2008}.  The beaming angular width is frequency-dependent, with higher frequencies subtending large opening angles, consistent with refraction in a dispersive medium \citep[e.g.,][]{Menietti:2011}.

The physical conditions in planetary magnetospheres are likely much different from stellar coronae, but the narrowness of the pulses suggests that stellar ECM radiation is also beamed. The beaming opening angle could also be frequency dependent, since the coronal plasma will be dispersive. Hence we include a frequency-dependent elliptical beaming weighting function for an ovoid hollow cone beam,

\begin{figure*}[h!]
\begin{center}
\includegraphics[width=7.in]{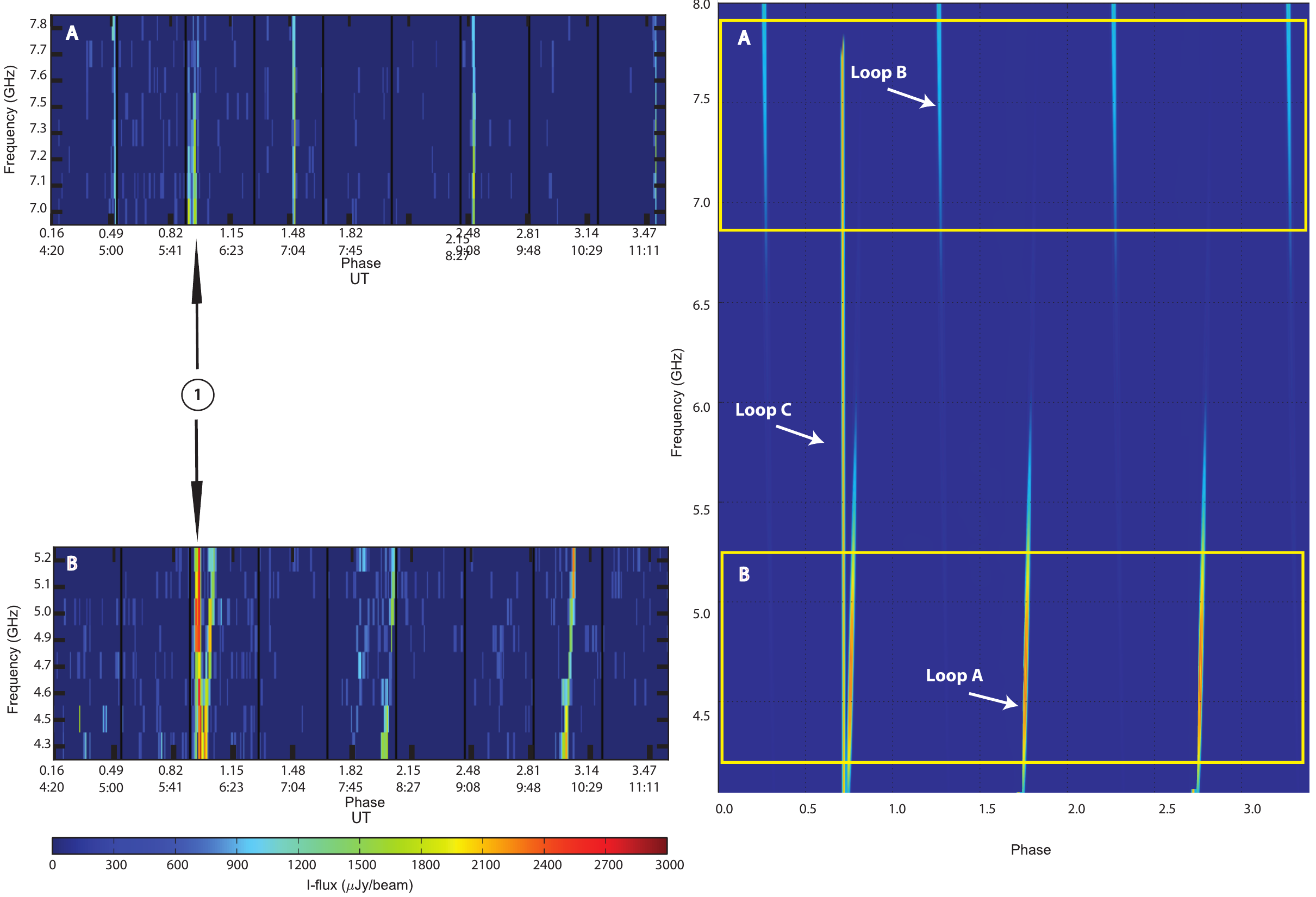}
\end{center}
\caption{({\it left}) Observed dynamic spectrum of  2M0746+20 on 16 Dec 2010 from 04:20 to 11:11 UT in the frequency bands 4.25--5.25~GHz and 6.95--7.85~GHz. All pulses are left-circularly polarized, except the pulse labeled 1, which is right-circularly polarized. This is shown in more detail in Figure \ref{fig:2m0746-0600-pulse}.  ({\it right}) Model dynamic spectrum from 4--8~GHz from rotating oblique coronal model using parameters in Table~\ref{table:cmi-model}. Loop labels refer to model loops for \MASS\ shown in Figure \ref{fig:3-d-views}.}
\label{fig:2m0746-obs-vs-model}
\end{figure*}

\begin{figure*}
\begin{center}
\includegraphics[width=7.0in]{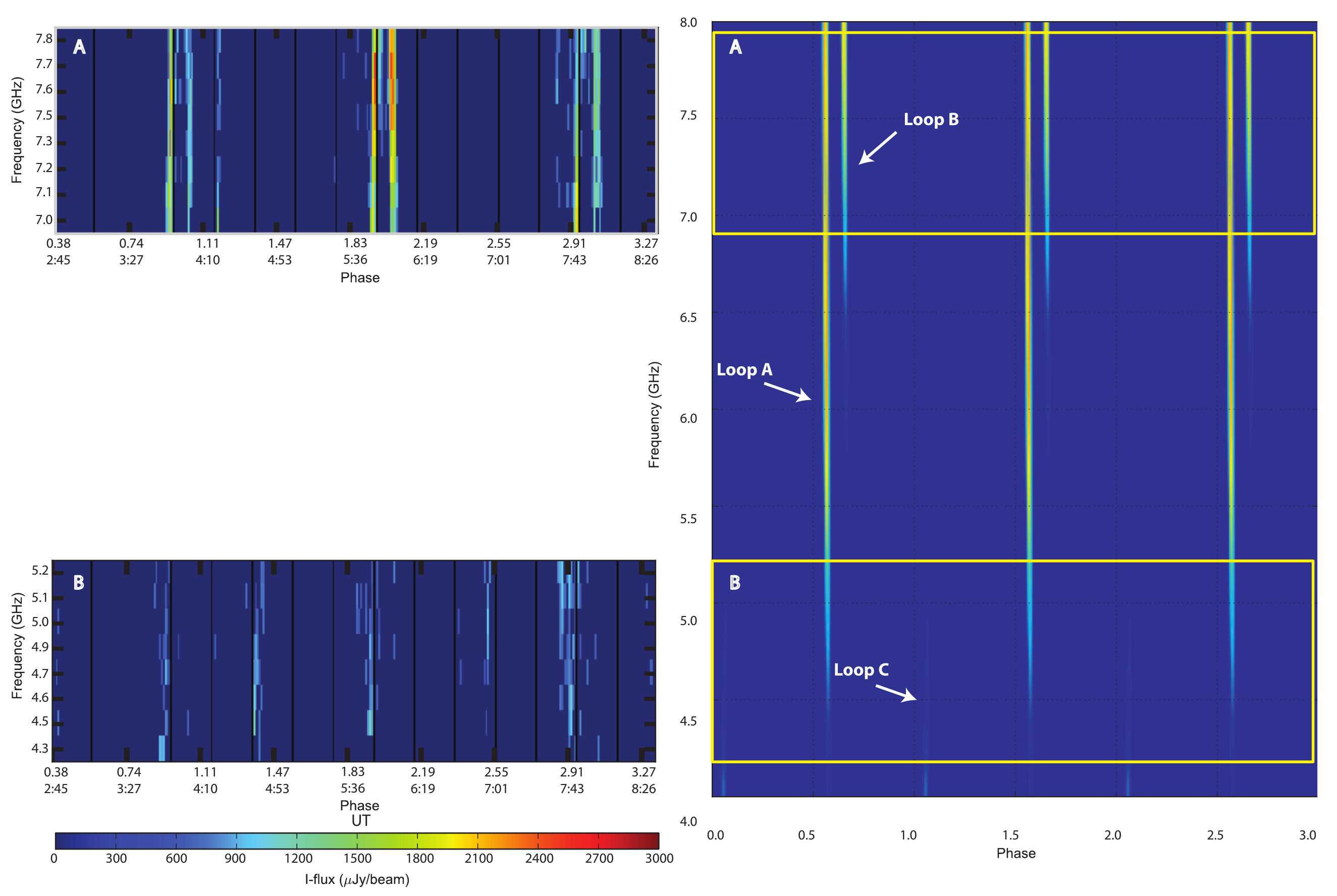}
\end{center}
\caption{({\it left}) Observed dynamic spectrum of  \TVLM\ on 2 July 2011 from 02:45 to 08:26 UT in the frequency bands 4.25--5.25~GHz and 6.95--7.85~GHz. ({\it right}) Model dynamical spectrum from 4--8~GHz from rotating oblique coronal model with parameters in Table~\ref{table:cmi-model}. Loop labels refer to model loops for \TVLM\ shown in Figure \ref{fig:3-d-views}.}
\label{fig:tvlm0513-obs-vs-model}
\end{figure*}

\begin{figure*}
\begin{center}
\includegraphics[width=6.5in]{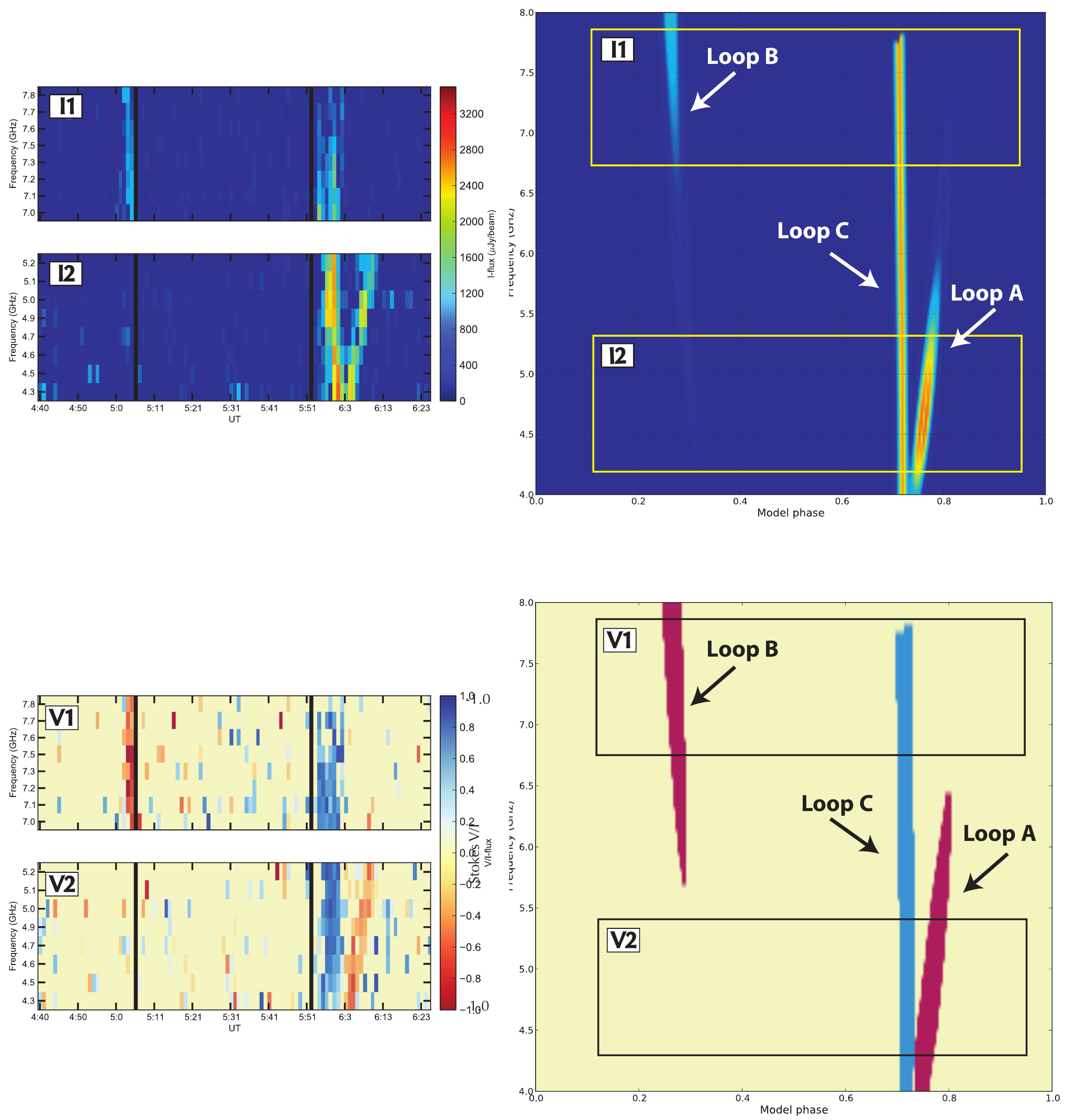}
\end{center}
\caption{Dynamic spectrum of  \MASS\  on 16 Dec 2010 from 04:40 to 06:25 UT in the frequency bands  6.95--7.85~GHz (I1: Stokes I, V1: Stokes V) and 4.25--5.25~GHz (I2: Stokes I, V2: Stokes V). The corresponding model dynamical spectra from 4--8~GHz from a rotating oblique coronal model with parameters given in Table~\ref{table:cmi-model} are shown in the right-hand panels. Loop labels refer to model loops for \MASS\ shown in Figure \ref{fig:3-d-views}.}
\label{fig:2m0746-0600-pulse}
\end{figure*}

\begin{figure*}
\begin{center}
\includegraphics[width=6.5in]{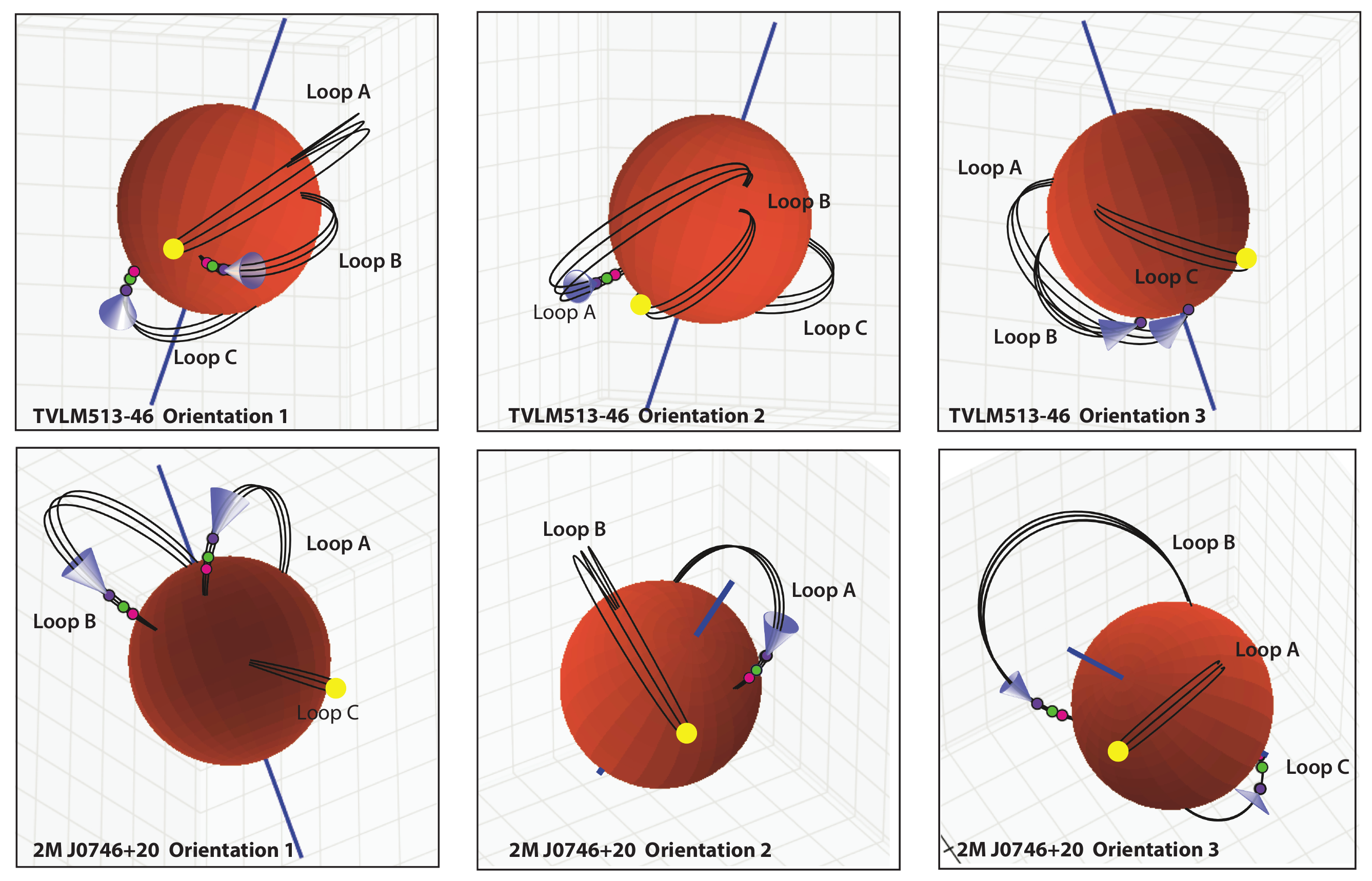}
\end{center}
\caption{Perspective views of coronal loop models for \TVLM\ (top) and \MASS\ (bottom) described by parameters in Table~\ref{table:cmi-model}. Location of ECM sources at 4~GHz (purple dots), 6~GHz (green dots), and 8~GHz (red dots) are shown at the  locations on the coronal loop corresponding to the local electron gyro-frequencies. The locations of sources beamed toward the observer are shown with yellow dots. }
\label{fig:3-d-views}
\end{figure*}

\begin{deluxetable*}{l c l l l l l l l l l l l l l}[h!]
\tablecolumns{14}
\tablecaption{ ECM pulsed emission model parameters}
\tablehead
{
 & & \multicolumn{5}{c}{Geometrical parameters} &  & &
\multicolumn{5}{c}{Beaming parameters} \\
\colhead{Star} & 
\colhead{Loop} &
\colhead{$\theta_{los}$} & 
\colhead{$\theta_{B}$} &
\colhead{$B_{0}$} &
\colhead{L}  & 
\colhead{N,S} &
\colhead{$\phi_m$} &  
\hspace{0.2in} &
\colhead{$\nu_0$} &
\colhead{$\theta_0$} &
\colhead{$\sigma_{\theta} $ } &
\colhead{$\phi   $ } &
\colhead{$\sigma_{\phi}$ }&
}
\startdata
\MASS & A & 36\degr & 80\degr & 2.5 kG & 1.7 & N & 150\degr & & 5~GHz  &1\degr   & 2\degr & 180\degr & 45\degr  \\
\MASS & B & 36\degr & 85\degr & 2.5 kG & 2.5 & N & 210\degr & & 5 GHz  &1\degr   & 2\degr & 180\degr & 45\degr \\
\MASS & C & 36\degr & 70\degr & 2.2 kG & 1.25 & S & 60\degr & & 5 GHz  &80\degr   & 2\degr & 180\degr & 45\degr \\ \\
\TVLM & A & 70\degr & 35\degr & 2.5 kG & 2.7 & N &  270\degr & & 5 GHz  &2\degr & 3\degr & 180\degr  & 45\degr  \\
\TVLM & B & 70\degr & 40\degr & 2.5 kG & 1.7 & N & 245\degr & & 5 GHz  &2\degr & 3\degr & 180\degr   & 45\degr  \\
\TVLM  & C & 70\degr & 60\degr & 2.5 kG & 2.0 & N & 120\degr & & 5 GHz  &2\degr & 3\degr & 180\degr  & 45\degr   
\enddata
\label{table:cmi-model}
\end{deluxetable*}

\begin{equation}
W(\theta ,\varphi ) = \exp \left\{ { - {{\left( {\frac{{\theta  - {\theta _c}}}{{{\sigma _\theta }}}} \right)}^2}} \right\}\exp \left\{ { - {{\left( {\frac{\phi }{{{\sigma _\phi }}}} \right)}^2}} \right\},
\end{equation}

where angles $(\theta, \sigma_{\theta})$ are the emission cone opening angle and width measured with respect to the magnetic field tangent direction at the source,  $(\phi, \sigma_{\phi})$ are the tangent plane angle and width, measured in a plane normal to the magnetic field, with $\phi=0\degr$ along the direction given by the cross product of the tangent vector and a vector pointing toward the magnetic pole. The frequency scaling  of the cone opening angle is given by
\begin{equation}
{\theta _c}\left( \nu  \right) = {\theta_0}{\left( {\frac{\nu }{{{\nu _0}}}} \right)^\beta }.
\end{equation}

The parameters $\theta_0$, $\nu_0$, $\beta$,  $\sigma_{\theta}$, $\theta_c$, and $\sigma_{\phi}$ are adjusted for best-fit to the observed dynamic spectra. In addition, we varied the rotation axis inclination angle ($\theta_{los}$), the inclination of the magnetic field loop with respect to the rotation axis ($\theta_B$), the loop extent (L-shell value), and the magnetic longitude ($\phi_m$) where $\phi_m = 0\degr$ when the magnetic axis is in the plane defined by the rotation axis and the observer's line of sight.

The dynamic spectra of both stars exhibited multiple pulses per period at most epochs (Figures~\ref{fig:2M0746-dynspectra}, \ref{fig:TVLM513-dynspectra}). We initially tried to model these pulses using a inclined dipolar magnetosphere with several `active' magnetic longitudes, similar to the model of \citet{Kuznetsov:2012}. We could not find any parameters for which the model dynamic spectra provided a satisfactory match to the observed dynamic spectra. We then tried a model using several localized active magnetic loops. Each loop is a subset of magnetic field lines associated with a magnetic dipole with prescribed surface field strength ($B_0$),  inclination to the rotation axis ($\theta_B$), active magnetic longitude ($\phi_m$), and loop size (magnetic L-shell value $L$).

We chose representative  dynamic spectra (16 Dec 2010 for \MASS, 2 July 2011 for \TVLM) to compare with model spectra. We then performed a systematic search in parameter space to find model spectra whose features best matched  the observed pulse spectra. These are shown in Figures~\ref{fig:2m0746-obs-vs-model} and \ref{fig:tvlm0513-obs-vs-model} respectively using parameters listed in Table~\ref{table:cmi-model}. The corresponding 3-dimensional perspective views of source locations and corresponding loop structures for each star are shown in Figure~\ref{fig:3-d-views}. 
Comparison between model  and observed spectra shows that the model reproduces all significant features of the observed pulses, including high and low frequency cutoffs, frequency drifting pulses, pulse-dependent circular polarization, and relative phasing of the pulses for each star.

\subsubsection{ECM Source locations}
The observed spectra for both stars were successfully modeled by ECM sources located on isolated active loops.  For \MASS\  Loop A, that produces low-frequency pulses below 6.5~GHz,  Loop B, that produces pulses above 6.5~GHz , and Loop C,  responsible for the isolated wideband burst of opposite circular polarization near 05:55 UT on 16 Dec 2010 (Figure 7). The loops have L-shell values of 1.7, 2.5, and 1.25  respectively,  and are well-separated in longitude.  

Note the large frequency drift ($df/dt\sim2$ MHz/sec) in the low-frequency pulse. This drift is a result of the narrow, tangent-directed (along $B$) beam and  frequency-dependent locations along the curved magnetic loop (viz. where the frequency is the local electron cyclotron frequency). As the star rotates, the frequency-dependent beam illuminates the observer at slightly different orientations which depend on loop location, and hence on frequency. 

\subsubsection{Rotation axis inclination angles}\label{sec:inclination angles}

For \TVLM, we ran model simulations with a range of inclination angles (i.e., between the star's rotation axis and the observer's line of sight) from 0\degr\ to 90\degr. The best-fit inclination was $i = $70\degr, although angles within 5\degr\ of this inclination also provided reasonable fits. This high inclination angle is consistent with the results of \citet{Hallinan:2008} who analyzed three UCD's with pulsed radio emission. They suggested that detection of pulsed radio emission is a geometrical selection effect: For all three UCD's, the measured \vsini and estimated radii implied rotation axes nearly perpendicular to the observer's line of sight. In particular, for \TVLM\  they found an inclination angle in the range $62.5\degr < i < 90\degr$, in good agreement with our model. 

By contrast, the inclination of  \MASS\  is not consistent with this scheme.
By combining long-term photometric variability analysis with \vsini observations, \citet{Harding:2013} determined a spin axis inclination $\theta_{los}=36\degr\pm4\degr$ for the secondary component, which is  the source of the pulsed radio emission. We  therefore have adopted this fixed value in the model, and found a good fit to the sample dynamic spectrum (cf. Figures~\ref{fig:2m0746-obs-vs-model}, \ref{fig:2m0746-0600-pulse}). However, we  consider this agreement a consistency check of the inclination rather than confirmation, since it is not clear whether other inclinations would also have provided an adequate fit.
   
\subsubsection{Beaming angles}

With one exception, the best-fit models for all sources on both stars required highly beamed emission in narrow cones ($1\degr-2\degr$) oriented tangent to the magnetic field direction at the source. The cones are so narrow that it was not possible to determine if they are filled or hollow. An exception was the intense pulse that appeared only once, on 16 Dec 2010 at 06:00 UT, and was oppositely-polarized to the main pulse at this epoch. The beaming angle for this pulse was best-fit using an opening angle $\theta_0 = 80\degr$. The narrow width of the pulse required a hollow cone with a correspondingly narrow thickness (2\degr). A close-up of this episodic pulse is shown in Figure~\ref{fig:2m0746-0600-pulse}. There is excellent agreement between the model and observed pulse frequency drift, and the frequency extent of both oppositely polarized pulses. This nearly perpendicular ECM beaming angle is commonly seen for ECM emission at both Earth (Ergun et al. 1998) and Jupiter (Imai et al. 2008). 

However, most pulses appear to be narrowly beamed along the magnetic field direction. What causes the narrow beaming along the magnetic field direction? In standard ECM theory, one expects maximum growth rate perpendicular to the magnetic field for shell-type electron distributions, or at an oblique angle for loss-cone distributions \citep{Treumann:2006,Mutel:2007}. For loss-cone driven CMI, the half-width angle of the radiation cone $\theta_b$ measured with respect to the B-field is given by \citep{Melrose:1982},

\begin{equation}
\theta_{b} = {\rm arccos}(v_{b}/c),
\end{equation}

where $v_b$ is the velocity of the electron beam. The model cones have half widths $\theta_0\sim1-2\degr$, which implies $0.9994c<v_b<0.9998c$ or electron beam energies $15<E<30$ MeV. 

These energies, although much larger than the beam energies exciting ECM emission on solar system planets, cannot be ruled out in a stellar coronal environment. Furthermore, with these high beam energies, the usual low-density condition for high ECM growth rates is no longer the case. More generally, we have \citep{Mutel:2006} 
\begin{equation}
\frac{\nu_{pe}}{\nu_{ce}} < {\left(\frac{\gamma -1}{\gamma}\right)}^{0.5},
\end{equation}
where $\nu_{pe}, \nu_{ce}$ are the electron plasma and cyclotron frequencies, and $\gamma$ is the beam Lorentz factor. For $\gamma \ll 1$ we recover the usual low-density constraint  ($\nu_{pe} \ll \nu_{ce}$), but for $\gamma \gg 1$, the right-hand side is $\sim$1. In this case, the inequality can be recast in terms of the electron density,
\begin{equation}
n_e < 10^{11} \cdot B_{kG}\ \ \text{cm}^{-3},
\end{equation}
where $B_{kG}$ is the magnetic field in kilogauss. These high densities may allow CMI growth very near the stellar photosphere.

However, the growth rate for loss-cone distributions diminishes rapidly with increasing loss-cone angle \citep{Pritchett:1986, Treumann:2006}, so a detailed calculation for these conditions is needed to determine if there is sufficient growth. In particular, \cite{Menietti:2011} found that  AKR dynamic `V shaped' spectra are well-fit by CMI emission extending several thousand km along a single active field line. Strong refraction, possibly resulting from sharp density walls near the ECM sources, can produce upward ray bending, as seen in terrestrial AKR beaming studies \citep{Ergun:1998, Mutel:2008}. However,  the extreme narrowness of the beamed emission in these stars is unlike any planetary ECM beaming, and may point to  a much denser refracting cavity, or perhaps a fundamentally different emission process.

\subsubsection{Model uniqueness and parameter uncertainties}

We computed model dynamic spectra for more than 50,000 sets of model parameters, choosing the model dynamic spectra that best-fit the observed profiles by visually comparing model and observed dynamic spectra.  The parameter search could not span all parameter space, since the total number of possible trials ($11^n$, where $10<n<36$ depending on the parameter) would have been computational prohibitive. Although the model and observed spectra agree very well, it is possible other source locations and beaming parameters could fit the observed spectra. Hence,  the models presented here may not be the {\it only} possible locations of ECM sources at the time of the observations. Nevertheless, in large measure they reproduce the frequency and phase dependence of the observed pulse emission, and hence are a tenable representation of the physical locations of the pulsed emission. 

What are the uncertainties in each of the best-fit parameters listed in Table~\ref{table:cmi-model}? Since the parameter search scheme comprised visual comparison of observed and model spectra, a formal calculation of parameter uncertainty is not applicable. However, the observed and model pulses differed significantly for angular differences ($\theta_{los}$, $\theta_B$, $\phi_m$) more than 5\degr\ from the best-fit values, L-shell variations more than 20\%, and beaming parameters $\theta_0$, $\sigma_{\theta}$ more than a few degrees. The beaming ovoid  parameters ($\phi$, $\sigma_{\phi}$) were less-well constrained, with significant model-observation differences only seen with angular changes exceeding about 30\degr. 

\subsubsection{Implications of magnetic field topology}
We were unable to find source locations that reproduced the observed pulses using a single dipolar magnetosphere with a set of `active' longitudes. Rather, the best-fit models comprise several active loops oriented at arbitrary inclinations with respect to each other. Although the model parameters may not be unique, they are consistent with the view that both stars' magnetic geometry consist of a small number of isolated regions with strong (kG) magnetic fields, rather than an overall strong field with azimuthal symmetry, or possibly a high-order multipole field. This picture supports recent suggestions \citep{Morin:2010,Williams:2014, Cook:2013} that radio-loud UCD's have  overall weak, non-axisymmetric magnetic fields with localized regions of stronger field strength.

\subsection{Quiescent Emission: Gyrosynchrotron Model}\label{sec:gs-model}
\begin{figure}[t!]
\begin{center}
\includegraphics[width=3.5in]{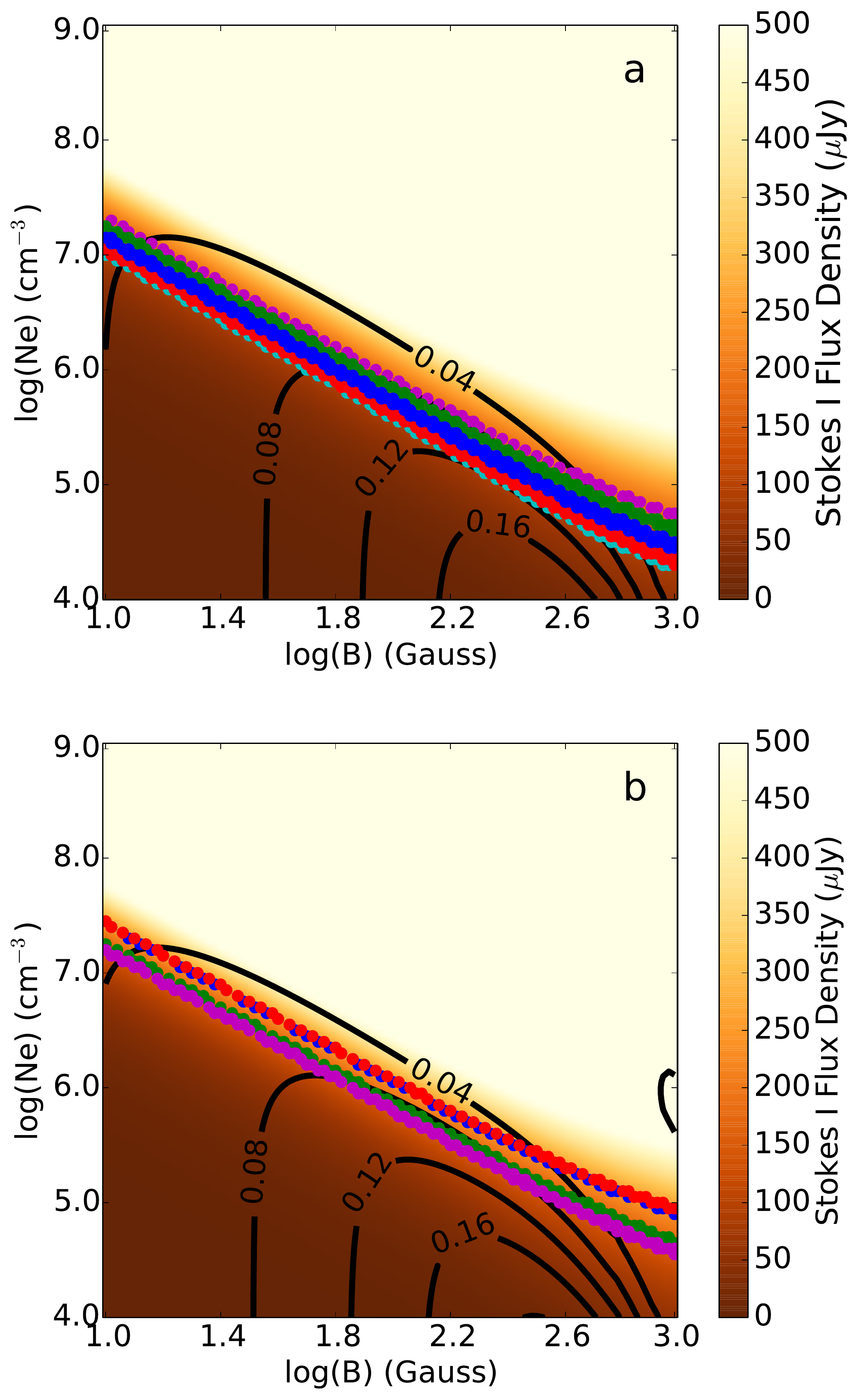}
\end{center}
\caption{Comparison between the observed and modeled 5~GHz flux density (color) and fractional circular polarization (black contours) for:  (a)  \MASS; (b) \TVLM. The model values are calculated using a gyrosynchrotron emission model for a range of surface magnetic field strengths and relativistic electron densities. The different color lines indicate regions where the model integrated flux density and circular polarization agree with measured values taking into account the measurements' uncertainty. The colors correspond to a different observing epoch as described in Figure \ref{fig:GSModel}.}
\label{fig:2dSurface}
\end{figure}

The observed spectral indices,  modest fractional circular polarizations, and inferred brightness temperatures (section \ref{sec:quiescent-obs}) all support the hypothesis that gyrosynchrotron radiation is responsible for the quiescent radio emission. In order to characterize the plasma conditions responsible for this emission, we constructed a model with mildly relativistic power-law electron populations immersed in a dipole magnetic field.

In order to speed up the numerical calculations, instead of calculating the $k$$\cdot$$B$ angle at every point on each line of sight, we used a weighted average of this angle along an effective path length $R$ determined by the relativistic electron density, which was assumed to have a quadratic dependence with radial distance ($n_e \propto r^{-2}$). There is no observational constraint on the distribution of electrons in the magnetospheres of UCDs. However, this distribution is necessary if we are to model the emission from these sources. We choose an inverse square electron density because this distribution is known to exist for the Sun at large distances. Additionally we ran the gyrosynchrotron model using a constant non-thermal electron density distribution and found the range of acceptable physical parameters was shifted by only 0.3 dex and does not significantly change our results.
 
We integrated the equation of radiative transfer along lines of sight distributed on a uniformly spaced grid in radius and azimuthal angle, where the axis of symmetry was the line between the observer and the star center. The lines of sight were spaced in intervals of 0.1 stellar radii from 1 to 3 stellar radii, and by 15\degr\ in azimuthal angle. We used the numerical expressions of \citet{Dulk:1985} for the absorption and emission coefficients for gyrosynchrotron radiation from a power-law electron distribution. Stokes  I and V fluxes were calculated by summing and differencing the extraordinary and ordinary mode emission over all lines of sight. 

We calculated model values for 5.0~GHz emission using a large range of relativistic electron densities,  $N_e$, and surface magnetic fields, $B$.  We assumed an average value for the energy power-law index for the relativistic electrons. Using the values in table \ref{table:QE}, the average spectral index for \MASS\ is -0.47$\pm$0.1 and for \TVLM\ -0.36$\pm$0.04.  Assuming the observed quiescent emission is optically thin, the energy power-law index is then 1.88$\pm$0.1 for \MASS\ and 1.76$\pm$0.04 for \TVLM.

The magnetic loop geometry for the modeled ECM emission is only applicable to the localized field at the source region and does not describe a global field. Thus in our gyrosynchrotron model we can choose any orientation for the global magnetic field. We assumed the inclination of the rotation axis, $\theta_{los}$, listed in table \ref{table:cmi-model} and varied the inclination of the magnetic axis with respect to the rotation axis, $\theta_B$, between 0-90$^{\circ}$. For all magnetic orientations, we found that the modeled fractional circular polarization was always in agreement with the measured upper limits and the modeled total flux does not change significantly. Thus our observations do not place constraints on the geometry of the global magnetic field.

Figure \ref{fig:2dSurface} shows a comparison between the observed and modeled total flux density (color) and circular polarization (black contours) in this parameter space. The different color lines in this Figure indicate regions where the modeled and measured integrated flux density are in agreement, taking into account the measurement's uncertainty. The colors correspond to different observing epochs as designated in Figure \ref{fig:GSModel}. It is clear from the comparison in Figure \ref{fig:2dSurface} that there is a degeneracy between values of $N_e$ and $B$, such that measured flux density values lie on lines of constant slope: 
 \begin{equation}
 \frac{\log(N_e)}{\log(B)}\sim\ -1.4
\end{equation} 
Thus for a given range of model values in agreement with the measured values, constraining the surface field constrains the corresponding relativistic electron density and vice versa. This degeneracy is broken if the fractional circular polarization can be measured. However these observations do not have sufficient signal-to-noise to make such measurements and the upper limits given in section \ref{sec:GS-Pol} are too high to place any constraint on the surface magnetic field strength.

\begin{figure}[t!]
\begin{center}
\includegraphics[width=3.5in]{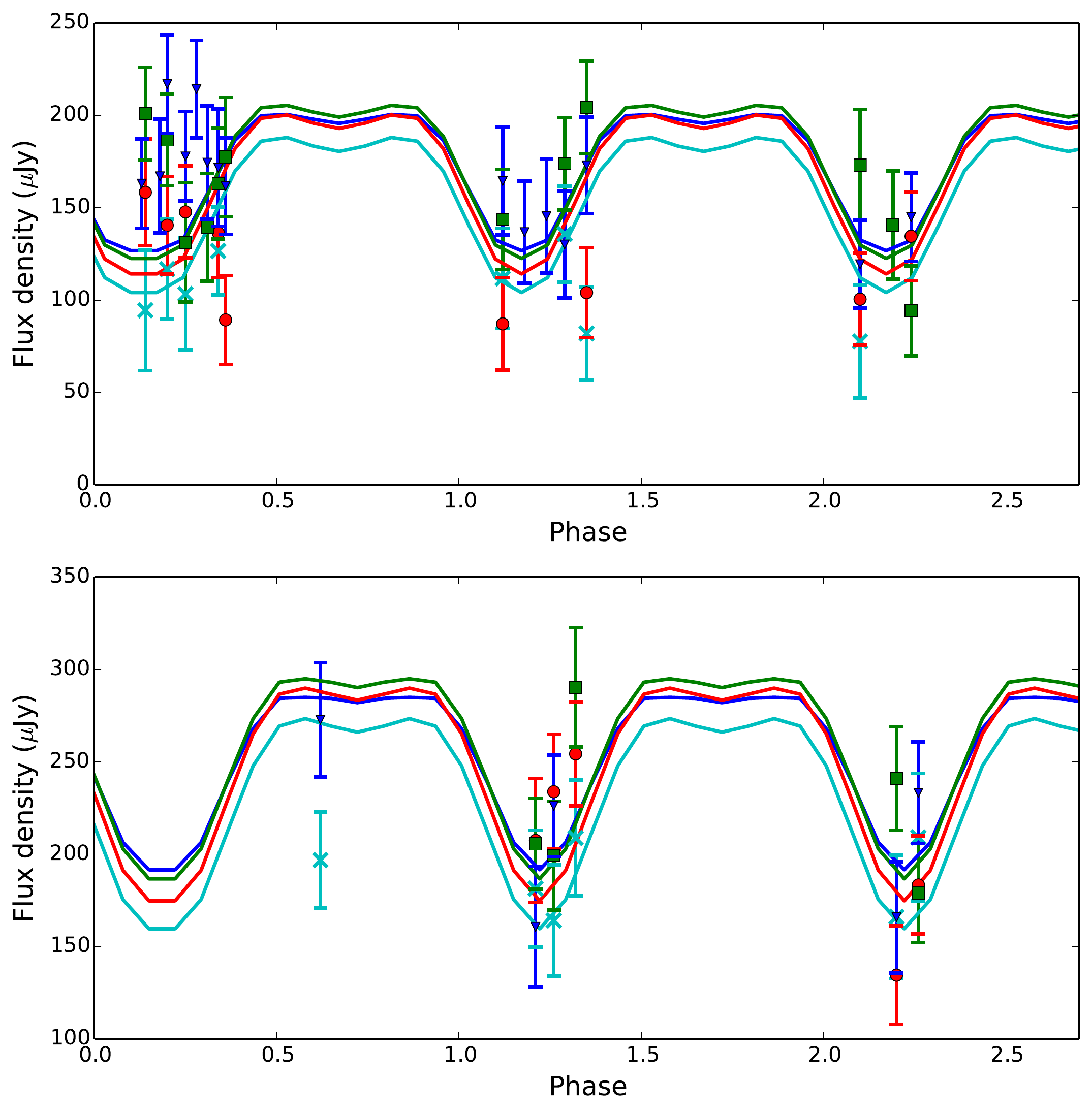}
\end{center}
\caption{(a)  \MASS\ Stokes I flux densities from  04:20$-$11:20 UT on 16 December 2010. The abscissa has been converted to phase using the ephemeris in section \ref{sec:pulse-model}. Each point represents the mean flux density in a 7 minute interval with 512~MHz bandwidth centered at 4.5~GHz (blue dots), 5.0~GHz (green dots), 7.1~GHz (red dots), and 7.7~GHz (cyan dots). Model light curves with the same parameters as the model SED in Figure \ref{fig:GSModel} are shown in solid lines with the same frequency-color coding. (b) Same as (a), but for \TVLM\ on 7 May 2011 from 06:25$-$12:38 UT. Note that the quiescent emission could not be sampled during pulsed emission  intervals.}
\label{fig:quiescent light curves}
\end{figure}

Comparison of the observed and model light curves (Figure \ref{fig:quiescent light curves}), although consistent within the data uncertainties, is less satisfactory. This is primarily because the the data uncertainties are comparable with or larger than the  predicted flux variations with phase. In addition, the largest model flux variations are during phases that were dominated by pulsed mission, which was excised to prevent contamination of the continuum emission. Additional observations at higher frequencies ($\nu > 10$~GHz) would be useful in studying the quiescent component, since the ECM radio pulses are unlikely  to occur at these frequencies. 

\section{Conclusions}

We performed multi-epoch, wide-band radio observations of two ultra-cool dwarfs, \MASS\ and \TVLM\ using the VLA. By combining our observations with archival data on the same objects spanning several years, we solved for new pulse periods, and modeled the dynamic spectra of both pulsed and continuum emission to derive physical conditions and source locations in the stellar magnetospheres. 

1. The dynamic spectra of the pulsed emission from both stars exhibited complex, time-variable morphologies, including high and low frequency cutoffs, frequency drifts, and variable circular polarization (ranging from less than 20\% to 100\%).

2. The angle between the electron cyclotron maser (ECM) emission cone direction and the magnetic field direction at the source is much smaller than ECM beaming observed in planetary magnetospheres. This may indicate that the stellar ECM sources are driven by a loss-cone instability with high electron beam energies ($15-30$ MeV). Alternatively, the sources could reside in  density  cavities with much sharper boundaries that refract the radiation sharply upward along the magnetic field.

4. Analysis of pulse timings over several years indicate stable periods, but long time intervals between sampling epochs allow the possibility of phase shifts in the gaps. Hence it is unclear whether individual active regions persist over several weeks, or over much longer periods (years). This can only be resolved with adequate temporal sampling.
The pulse period of \MASS\ is compatible with previously published values, but is more precise, indicating a stable period within 0.2 sec over several years.   

5.  We were able to model the pulsed emission by ECM radiation from a small number of loops of high magnetic field (2-3 kG) with radial extents $\sim1.2-2.7$ stellar radii. The loops are well-separated in magnetic longitude, and are not part of a single dipolar magnetosphere. However, we did not test the possibility that the emission could be modeled using higher order multipolar fields and cannot rule out this field morphology.

For \TVLM\ the best-fit model  has a high rotation axis inclination to the observer's line of sight, consistent with previous modeling \citep{Kuznetsov:2012}. For \MASS, the lower inclination reported by \citet{Harding:2013} was used in the model. This provides a good fit to the observed dynamic spectra, but we cannot rule out models with higher inclinations. 

6.  The continuous radiation is well-modeled by gyrosynchrotron emission from mildly relativistic power-law electrons in a dipolar magnetic field. However, in this model there is a clear degeneracy between the electron density and surface magnetic field strength that is unbroken by the presented observations.

Considering the last two results, the overall magnetic configuration of both stars support recent suggestions that radio over-luminous UCD's have 'weak field' non-axisymmetric topologies \citep{Cook:2013,Williams:2014}, but with isolated regions of high magnetic field.

\vspace{0.2in}
We thank Evan Abbuhl with help preparing the illustrations. We are grateful for financial support  from the National Science Foundation through research grant AST-0908941. This research has made use of the SIMBAD database, operated at CDS, Strasbourg, France.

Facilities: Karl G. Jansky VLA

\bibliography{ucd-papers}

\begin{thebibliography}{64}
\expandafter\ifx\csname natexlab\endcsname\relax\def\natexlab#1{#1}\fi

\bibitem[{{Allred} {et~al.}(2006){Allred}, {Hawley}, {Abbett}, \&
  {Carlsson}}]{Allred:2006}
{Allred}, J.~C., {Hawley}, S.~L., {Abbett}, W.~P., \& {Carlsson}, M. 2006,
  \apj, 644, 484

\bibitem[{{Antonova} {et~al.}(2008){Antonova}, {Doyle}, {Hallinan}, {Bourke},
  \& {Golden}}]{Antonova:2008}
{Antonova}, A., {Doyle}, J.~G., {Hallinan}, G., {Bourke}, S., \& {Golden}, A.
  2008, \aap, 487, 317

\bibitem[{{Antonova} {et~al.}(2013){Antonova}, {Hallinan}, {Doyle}, {Yu},
  {Kuznetsov}, {Metodieva}, {Golden}, \& {Cruz}}]{Antonova:2013}
{Antonova}, A., {Hallinan}, G., {Doyle}, J.~G., {Yu}, S., {Kuznetsov}, A.,
  {Metodieva}, Y., {Golden}, A., \& {Cruz}, K.~L. 2013, \aap, 549, A131

\bibitem[{Berger(2002)}]{Berger:2002}
Berger, E. 2002, \apj, 572, 503

\bibitem[{Berger(2006)}]{Berger:2006}
---. 2006, \apj, 648, 629

\bibitem[{Berger {et~al.}(2001)Berger, Ball, Becker, Clarke, Frail, Fukuda,
  Hoffman, Mellon, Momjian, Murphy, Teng, Woodruff, Zauderer, \&
  Zavala}]{Berger:2001}
Berger, E., Ball, S., Becker, K.~M., Clarke, M., Frail, D.~A., Fukuda, T.~A.,
  Hoffman, I.~M., Mellon, R., Momjian, E., Murphy, N.~W., Teng, S.~H.,
  Woodruff, T., Zauderer, B.~A., \& Zavala, R.~T. 2001, Nature, 410, 338

\bibitem[{Berger {et~al.}(2008)Berger, Gizis, Giampapa, Rutledge, Liebert,
  Mart{\'\i}n, Basri, Fleming, Johns-Krull, Phan-Bao, \& Sherry}]{Berger:2008a}
Berger, E., Gizis, J.~E., Giampapa, M.~S., Rutledge, R.~E., Liebert, J.,
  Mart{\'\i}n, E., Basri, G., Fleming, T.~A., Johns-Krull, C.~M., Phan-Bao, N.,
  \& Sherry, W.~H. 2008, The Astrophysical Journal, 673, 1080

\bibitem[{Berger {et~al.}(2009)Berger, Rutledge, Phan-Bao, Basri, Giampapa,
  Gizis, Liebert, Mart{\'\i}n, \& Fleming}]{Berger:2009}
Berger, E., Rutledge, R.~E., Phan-Bao, N., Basri, G., Giampapa, M.~S., Gizis,
  J.~E., Liebert, J., Mart{\'\i}n, E., \& Fleming, T.~A. 2009, The
  Astrophysical Journal, 695, 310

\bibitem[{{Browning}(2008)}]{Browning:2008}
{Browning}, M.~K. 2008, \apj, 676, 1262

\bibitem[{{Burgasser} \& {Putman}(2005)}]{Burgasser:2005}
{Burgasser}, A.~J., \& {Putman}, M.~E. 2005, \apj, 626, 486

\bibitem[{{Cook} {et~al.}(2013){Cook}, {Williams}, \& {Berger}}]{Cook:2013}
{Cook}, B.~A., {Williams}, P.~K.~G., \& {Berger}, E. 2013, ArXiv e-prints

\bibitem[{{Dahn} {et~al.}(2002){Dahn}, {Harris}, {Vrba}, {Guetter}, {Canzian},
  {Henden}, {Levine}, {Luginbuhl}, {Monet}, {Monet}, {Pier}, {Stone}, {Walker},
  {Burgasser}, {Gizis}, {Kirkpatrick}, {Liebert}, \& {Reid}}]{Dahn:2002}
{Dahn}, C.~C., {Harris}, H.~C., {Vrba}, F.~J., {Guetter}, H.~H., {Canzian}, B.,
  {Henden}, A.~A., {Levine}, S.~E., {Luginbuhl}, C.~B., {Monet}, A.~K.~B.,
  {Monet}, D.~G., {Pier}, J.~R., {Stone}, R.~C., {Walker}, R.~L., {Burgasser},
  A.~J., {Gizis}, J.~E., {Kirkpatrick}, J.~D., {Liebert}, J., \& {Reid}, I.~N.
  2002, Astronomical Journal, 124, 1170

\bibitem[{{Doyle} {et~al.}(2010){Doyle}, {Antonova}, {Marsh}, {Hallinan}, {Yu},
  \& {Golden}}]{Doyle:2010}
{Doyle}, J.~G., {Antonova}, A., {Marsh}, M.~S., {Hallinan}, G., {Yu}, S., \&
  {Golden}, A. 2010, \aap, 524, A15

\bibitem[{{Dulk}(1985)}]{Dulk:1985}
{Dulk}, G.~A. 1985, \araa, 23, 169

\bibitem[{{Durney} {et~al.}(1993){Durney}, {De Young}, \&
  {Roxburgh}}]{Durney:1993}
{Durney}, B.~R., {De Young}, D.~S., \& {Roxburgh}, I.~W. 1993, \solphys, 145,
  207

\bibitem[{Ergun {et~al.}(2000)Ergun, Carlson, McFadden, Delory, Strangeway, \&
  Pritchett}]{Ergun:2000}
Ergun, R.~E., Carlson, C.~W., McFadden, J.~P., Delory, G.~T., Strangeway,
  R.~J., \& Pritchett, P.~L. 2000, The Astrophysical Journal, 538, 456

\bibitem[{{Ergun} {et~al.}(1998){Ergun}, {Carlson}, {McFadden}, {Mozer},
  {Delory}, {Peria}, {Chaston}, {Temerin}, {Elphic}, {Strangeway}, {Pfaff},
  {Cattell}, {Klumpar}, {Shelley}, {Peterson}, {Moebius}, \&
  {Kistler}}]{Ergun:1998}
{Ergun}, R.~E., {Carlson}, C.~W., {McFadden}, J.~P., {Mozer}, F.~S., {Delory},
  G.~T., {Peria}, W., {Chaston}, C.~C., {Temerin}, M., {Elphic}, R.,
  {Strangeway}, R., {Pfaff}, R., {Cattell}, C.~A., {Klumpar}, D., {Shelley},
  E., {Peterson}, W., {Moebius}, E., \& {Kistler}, L. 1998, \grl, 25, 2061

\bibitem[{{Gastine} {et~al.}(2013){Gastine}, {Morin}, {Duarte}, {Reiners},
  {Christensen}, \& {Wicht}}]{Gastine:2013}
{Gastine}, T., {Morin}, J., {Duarte}, L., {Reiners}, A., {Christensen}, U.~R.,
  \& {Wicht}, J. 2013, \aap, 549, L5

\bibitem[{{Gizis} {et~al.}(2000){Gizis}, {Monet}, {Reid}, {Kirkpatrick},
  {Liebert}, \& {Williams}}]{Gizis:2000}
{Gizis}, J.~E., {Monet}, D.~G., {Reid}, I.~N., {Kirkpatrick}, J.~D., {Liebert},
  J., \& {Williams}, R.~J. 2000, \aj, 120, 1085

\bibitem[{G\"udel \& Benz(1993)}]{Gudel:1993a}
G\"udel, M., \& Benz, A.~O. 1993, Astrophysical Journal, 405, L63

\bibitem[{{Gurnett}(1974)}]{Gurnett:1974}
{Gurnett}, D.~A. 1974, \jgr, 79, 4227

\bibitem[{Hallinan {et~al.}(2006)Hallinan, Antonova, Doyle, Bourke, Brisken, \&
  Golden}]{Hallinan:2006}
Hallinan, G., Antonova, A., Doyle, J.~G., Bourke, S., Brisken, W.~F., \&
  Golden, A. 2006, \apj, 653, 690

\bibitem[{Hallinan {et~al.}(2008)Hallinan, Antonova, Doyle, Bourke, Lane, \&
  Golden}]{Hallinan:2008}
Hallinan, G., Antonova, A., Doyle, J.~G., Bourke, S., Lane, C., \& Golden, A.
  2008, arXiv.org, astro-ph

\bibitem[{Hallinan {et~al.}(2007)Hallinan, Bourke, Lane, Antonova, Zavala,
  Brisken, Boyle, Vrba, Doyle, \& Golden}]{Hallinan:2007}
Hallinan, G., Bourke, S., Lane, C., Antonova, A., Zavala, R.~T., Brisken,
  W.~F., Boyle, R.~P., Vrba, F.~J., Doyle, J.~G., \& Golden, A. 2007, \apj,
  663, L25

\bibitem[{{Harding} {et~al.}(2013{\natexlab{a}}){Harding}, {Hallinan}, {Boyle},
  {Golden}, {Singh}, {Sheehan}, {Zavala}, \& {Butler}}]{Harding:2013a}
{Harding}, L.~K., {Hallinan}, G., {Boyle}, R.~P., {Golden}, A., {Singh}, N.,
  {Sheehan}, B., {Zavala}, R.~T., \& {Butler}, R.~F. 2013{\natexlab{a}},
  Astrophysical Journal, 779, 101

\bibitem[{{Harding} {et~al.}(2013{\natexlab{b}}){Harding}, {Hallinan},
  {Konopacky}, {Kratter}, {Boyle}, {Butler}, \& {Golden}}]{Harding:2013}
{Harding}, L.~K., {Hallinan}, G., {Konopacky}, Q.~M., {Kratter}, K.~M.,
  {Boyle}, R.~P., {Butler}, R.~F., \& {Golden}, A. 2013{\natexlab{b}}, Astron.
  Astrophys., 554, A113

\bibitem[{Imai {et~al.}(2008)Imai, Imai, Higgins, \& Thieman}]{Imai:2008}
Imai, M., Imai, K., Higgins, C.~A., \& Thieman, J.~R. 2008, Geophysical
  Research Letters, 35

\bibitem[{{Konopacky} {et~al.}(2010){Konopacky}, {Ghez}, {Barman}, {Rice},
  {Bailey}, {White}, {McLean}, \& {Duch{\^e}ne}}]{Konopacky:2010}
{Konopacky}, Q.~M., {Ghez}, A.~M., {Barman}, T.~S., {Rice}, E.~L., {Bailey},
  III, J.~I., {White}, R.~J., {McLean}, I.~S., \& {Duch{\^e}ne}, G. 2010,
  Astrophysical Journal, 711, 1087

\bibitem[{{Konopacky} {et~al.}(2012){Konopacky}, {Ghez}, {Fabrycky},
  {Macintosh}, {White}, {Barman}, {Rice}, {Hallinan}, \&
  {Duch{\^e}ne}}]{Konopacky:2012}
{Konopacky}, Q.~M., {Ghez}, A.~M., {Fabrycky}, D.~C., {Macintosh}, B.~A.,
  {White}, R.~J., {Barman}, T.~S., {Rice}, E.~L., {Hallinan}, G., \&
  {Duch{\^e}ne}, G. 2012, Astrophysical Journal, 750, 79

\bibitem[{{Kuznetsov} {et~al.}(2012){Kuznetsov}, {Doyle}, {Yu}, {Hallinan},
  {Antonova}, \& {Golden}}]{Kuznetsov:2012}
{Kuznetsov}, A.~A., {Doyle}, J.~G., {Yu}, S., {Hallinan}, G., {Antonova}, A.,
  \& {Golden}, A. 2012, Astrophysical Journal, 746, 99

\bibitem[{{Lamy} {et~al.}(2011){Lamy}, {Cecconi}, {Zarka}, {Canu}, {Schippers},
  {Kurth}, {Mutel}, {Gurnett}, {Menietti}, \& {Louarn}}]{Lamy:2011}
{Lamy}, L., {Cecconi}, B., {Zarka}, P., {Canu}, P., {Schippers}, P., {Kurth},
  W.~S., {Mutel}, R.~L., {Gurnett}, D.~A., {Menietti}, D., \& {Louarn}, P.
  2011, Journal of Geophysical Research (Space Physics), 116, 4212

\bibitem[{Lamy {et~al.}(2008)Lamy, Zarka, Cecconi, Prang{\'e}, Kurth, \&
  Gurnett}]{Lamy:2008}
Lamy, L., Zarka, P., Cecconi, B., Prang{\'e}, R., Kurth, W.~S., \& Gurnett,
  D.~A. 2008, Journal of Geophysical Research, 113, 07201

\bibitem[{{Lee} {et~al.}(2013){Lee}, {Yi}, {Lim}, {Kim}, {Seough}, \&
  {Yoon}}]{Lee:2013}
{Lee}, S.-Y., {Yi}, S., {Lim}, D., {Kim}, H.-E., {Seough}, J., \& {Yoon}, P.~H.
  2013, Journal of Geophysical Research (Space Physics), 118, 7036

\bibitem[{{Machado} {et~al.}(1980){Machado}, {Avrett}, {Vernazza}, \&
  {Noyes}}]{Machado:1980}
{Machado}, M.~E., {Avrett}, E.~H., {Vernazza}, J.~E., \& {Noyes}, R.~W. 1980,
  \apj, 242, 336

\bibitem[{McLean {et~al.}(2011)McLean, Berger, Irwin, Forbrich, \&
  Reiners}]{McLean:2011}
McLean, M., Berger, E., Irwin, J., Forbrich, J., \& Reiners, A. 2011, \apj,
  741, 1

\bibitem[{{McLean} {et~al.}(2012){McLean}, {Berger}, \&
  {Reiners}}]{Mclean:2012}
{McLean}, M., {Berger}, E., \& {Reiners}, A. 2012, \apj, 746, 23

\bibitem[{Melrose \& Dulk(1982)}]{Melrose:1982}
Melrose, D.~B., \& Dulk, G.~A. 1982, Astrophysical Journal, 259, 844

\bibitem[{{Menietti} {et~al.}(2011){Menietti}, {Mutel}, {Christopher},
  {Hutchinson}, \& {Sigwarth}}]{Menietti:2011}
{Menietti}, J.~D., {Mutel}, R.~L., {Christopher}, I.~W., {Hutchinson}, K.~A.,
  \& {Sigwarth}, J.~B. 2011, Journal of Geophysical Research (Space Physics),
  116, 12219

\bibitem[{{Morin} {et~al.}(2010){Morin}, {Donati}, {Petit}, {Delfosse},
  {Forveille}, \& {Jardine}}]{Morin:2010}
{Morin}, J., {Donati}, J.-F., {Petit}, P., {Delfosse}, X., {Forveille}, T., \&
  {Jardine}, M.~M. 2010, \mnras, 407, 2269

\bibitem[{{Morin} {et~al.}(2011){Morin}, Dormy, Schrinner, \&
  {Donati}}]{Morin:2011}
{Morin}, J., Dormy, E., Schrinner, M., \& {Donati}, J.-F. 2011, \mnras, 418,
  L133

\bibitem[{{Mutel} {et~al.}(2008){Mutel}, {Christopher}, \&
  {Pickett}}]{Mutel:2008}
{Mutel}, R.~L., {Christopher}, I.~W., \& {Pickett}, J.~S. 2008, \grl, 35, 7104

\bibitem[{Mutel {et~al.}(2006)Mutel, Menietti, Christopher, Gurnett, \&
  Cook}]{Mutel:2006}
Mutel, R.~L., Menietti, J.~D., Christopher, I.~W., Gurnett, D.~A., \& Cook,
  J.~M. 2006, Journal of Geophysical Research, 111, 10203

\bibitem[{Mutel {et~al.}(1998)Mutel, Molnar, Waltman, \& Ghigo}]{Mutel:1998}
Mutel, R.~L., Molnar, L.~A., Waltman, E.~B., \& Ghigo, F.~D. 1998,
  Astrophysical Journal, 507, 371

\bibitem[{{Mutel} {et~al.}(2007){Mutel}, {Peterson}, {Jaeger}, \&
  {Scudder}}]{Mutel:2007}
{Mutel}, R.~L., {Peterson}, W.~M., {Jaeger}, T.~R., \& {Scudder}, J.~D. 2007,
  Journal of Geophysical Research (Space Physics), 112

\bibitem[{{Neuh{\"a}user} {et~al.}(1999){Neuh{\"a}user}, {Brice{\~n}o},
  {Comer{\'o}n}, {Hearty}, {Mart{\'{\i}}n}, {Schmitt}, {Stelzer}, {Supper},
  {Voges}, \& {Zinnecker}}]{Neuhauser:1999}
{Neuh{\"a}user}, R., {Brice{\~n}o}, C., {Comer{\'o}n}, F., {Hearty}, T.,
  {Mart{\'{\i}}n}, E.~L., {Schmitt}, J.~H.~M.~M., {Stelzer}, B., {Supper}, R.,
  {Voges}, W., \& {Zinnecker}, H. 1999, \aap, 343, 883

\bibitem[{{Neupert}(1968)}]{Neupert:1968}
{Neupert}, W.~M. 1968, \apjl, 153, L59

\bibitem[{Osten {et~al.}(2006{\natexlab{a}})Osten, Hawley, Allred, Johns-Krull,
  Brown, \& Harper}]{Osten:2006}
Osten, R.~A., Hawley, S.~L., Allred, J., Johns-Krull, C.~M., Brown, A., \&
  Harper, G.~M. 2006{\natexlab{a}}, \apj, 647, 1349

\bibitem[{Osten {et~al.}(2006{\natexlab{b}})Osten, Hawley, Bastian, \&
  Reid}]{Osten:2006b}
Osten, R.~A., Hawley, S.~L., Bastian, T.~S., \& Reid, I.~N. 2006{\natexlab{b}},
  Astrophysical Journal, 637, 518

\bibitem[{Osten \& Jayawardhana(2006)}]{Osten:2006a}
Osten, R.~A., \& Jayawardhana, R. 2006, Astrophysical Journal, 644, L67

\bibitem[{{Osten} \& {Wolk}(2009)}]{Osten:2009}
{Osten}, R.~A., \& {Wolk}, S.~J. 2009, \apj, 691, 1128

\bibitem[{Phan-Bao {et~al.}(2007)Phan-Bao, Osten, Lim, Mart{\^O}{\o}Ωn, \&
  Ho}]{Phan-Bao:2007}
Phan-Bao, N., Osten, R.~A., Lim, J., Mart{\^O}{\o}Ωn, E.~L., \& Ho, P. T.~P.
  2007, \apj, 658, 553

\bibitem[{{Pritchett}(1986)}]{Pritchett:1986}
{Pritchett}, P.~L. 1986, Journal of Geophysical Research, 91, 13569

\bibitem[{{Raedler} {et~al.}(1990){Raedler}, {Wiedemann}, {Brandenburg},
  {Meinel}, \& {Tuominen}}]{Raedler:1990}
{Raedler}, K.-H., {Wiedemann}, E., {Brandenburg}, A., {Meinel}, R., \&
  {Tuominen}, I. 1990, \aap, 239, 413

\bibitem[{{Reid} {et~al.}(2001){Reid}, {Gizis}, {Kirkpatrick}, \&
  {Koerner}}]{Reid:2001}
{Reid}, I.~N., {Gizis}, J.~E., {Kirkpatrick}, J.~D., \& {Koerner}, D.~W. 2001,
  Astronomical Journal, 121, 489

\bibitem[{{Reiners} \& {Basri}(2007)}]{Reiners:2007}
{Reiners}, A., \& {Basri}, G. 2007, \apj, 656, 1121

\bibitem[{{Richards} {et~al.}(2012){Richards}, {Agafonov}, \&
  {Sharova}}]{Richards:2012}
{Richards}, M.~T., {Agafonov}, M.~I., \& {Sharova}, O.~I. 2012, Astrophysical
  Journal, 760, 8

\bibitem[{Richards {et~al.}(2003)Richards, Waltman, Ghigo, \&
  Richards}]{Richards:2003}
Richards, M.~T., Waltman, E.~B., Ghigo, F.~D., \& Richards, D. S.~P. 2003,
  Astrophysical Journal Supplement Series, 147, 337

\bibitem[{{Route} \& {Wolszczan}(2012)}]{Route:2012}
{Route}, M., \& {Wolszczan}, A. 2012, \apjl, 747, L22

\bibitem[{Treumann(2006)}]{Treumann:2006}
Treumann, R.~A. 2006, The Astronomy and Astrophysics Review, 13, 229

\bibitem[{{West} {et~al.}(2004){West}, {Hawley}, {Walkowicz}, {Covey},
  {Silvestri}, {Raymond}, {Harris}, {Munn}, {McGehee}, {Ivezi{\'c}}, \&
  {Brinkmann}}]{West:2004}
{West}, A.~A., {Hawley}, S.~L., {Walkowicz}, L.~M., {Covey}, K.~R.,
  {Silvestri}, N.~M., {Raymond}, S.~N., {Harris}, H.~C., {Munn}, J.~A.,
  {McGehee}, P.~M., {Ivezi{\'c}}, {\v Z}., \& {Brinkmann}, J. 2004,
  Astronomical Journal, 128, 426

\bibitem[{{Williams} {et~al.}(2014){Williams}, {Cook}, \&
  {Berger}}]{Williams:2014}
{Williams}, P.~K.~G., {Cook}, B.~A., \& {Berger}, E. 2014, \apj, 785, 9

\bibitem[{{Wolszczan} \& {Route}(2014)}]{Wolszczan:2014}
{Wolszczan}, A., \& {Route}, M. 2014, \apj, 788, 23

\bibitem[{Zarka(1998)}]{Zarka:1998}
Zarka, P. 1998, Journal of Geophysical Research, 103, 20159

\bibitem[{Zarka(2004)}]{Zarka:2004}
---. 2004, Advances in Space Research, 33, 2045

\end{thebibliography}

\end{document}